\documentclass[useAMS,usenatbib]{mn2e}
\pdfoutput=1

\usepackage{graphicx}
\usepackage{amsmath}
\usepackage{amssymb}
\usepackage{color}
\newcommand{\pd}[2]{\frac{\partial #1}{\partial #2}}
\newcommand{\av}[1]{\bar{#1}}
\newcommand{\HALF}{\frac{1}{2}}
\newcommand{\DS}{\displaystyle}
\renewcommand{\vec}[1]{\bmath{#1}}
\newcommand{\hvec}[1]{\hat{\vec{#1}}}

\title[The Kinked Jet of the Crab Nebula]{Modelling the Kinked Jet of the Crab Nebula}
\author[Mignone et al.]{
A. Mignone$^{1}$\thanks{E-mail:mignone@ph.unito.it (AM)},
E. Striani$^{2,3,4}$, M. Tavani$^{2,3,4}$ and A. Ferrari$^{1}$\\
$^{1}$ Dipartimento di Fisica Generale ``Amedeo Avogadro'' Universit\`a degli Studi di Torino, Via Pietro Giuria 1, 10125 Torino, Italy\\
$^{2}$ INAF/IAPS-Roma, I-00133 Roma, Italy  \\
$^{3}$ Dip. di Fisica, Univ. Tor Vergata, I-00133 Roma, Italy\\
$^{4}$ INFN Roma Tor Vergata, I-00133 Roma, Italy\\
}

\begin{document}

\pagerange{\pageref{firstpage}--\pageref{lastpage}} \pubyear{2013}

\maketitle

\label{firstpage}
\begin{abstract}
  We investigate the dynamical propagation of the South-East jet from the Crab pulsar interacting with supernova ejecta by means of three-dimensional relativistic MHD numerical simulations with the PLUTO code.
  The initial jet structure is set up from the inner regions of the Crab Nebula.
  We study the evolution of hot, relativistic hollow outflows initially carrying a purely azimuthal magnetic field.
  Our jet models are characterized by different choices of the outflow magnetization ($\sigma$ parameter) and the bulk Lorentz factor ($\gamma_{j}$).

  We show that the jet is heavily affected by the growth of current-driven kink instabilities causing considerable deflection throughout its propagation length.
  This behavior is partially stabilized by the combined action of larger flow velocities and/or reduced magnetic field strengths.
  We find that our best jet models are characterized by relatively large values of $\sigma$ ($\gtrsim 1$) and small values of $\gamma_{j}\simeq 2$.
  Our results are in good agreement with the recent X-ray (\textit{Chandra}) data of the Crab Nebula South-East jet indicating that the jet changes direction of propagation on a time scale of the order of few years.
  The 3D models presented here may have important implications in the investigation of particle acceleration in relativistic outflows.
\end{abstract}


\section{Introduction}
\label{sec:introduction}
%
%
%
%
%
%
%

Pulsars loose their rotational energy through a relativistic wind of waves and particles.
The interaction of these outflows with the surrounding ambient produces Pulsar Wind Nebulae (PWNe), observable from radio to $\gamma$-rays.
Pulsar Wind Nebulae often show a torus-jet structure (see, e.g., \cite{Kargaltsev} for a review, and the Chandra images of the Crab, Vela and B1509-58 Nebulae).
Several theoretical \citep{Lyubarsky2001, Lyubarsky2002, Petri} and numerical \citep{KL.2003, dZAB.2004, dZVAB.2006} studies attempted to explain and reproduce this structure.

The Crab Nebula is surely the most popular and studied PWN. It's powered by a pulsar with a very large spin-down luminosity, $L_{sd}= 5 \times 10^{38}$ erg/s, that is carried away by a relativistic and highly magnetized wind. According to estimates of the number of pairs emitted by the Crab Pulsar, the energy flux carried by the wind is dominated by the Poynting flux while close to the star, and dominated by the particles when close to the termination shock (Kennel $\&$ Coroniti 1984).
First theoretical \citep{K&C} and numerical \citep{KL.2003, dZAB.2004} studies suggested that the magnetization parameter, defined as the ratio between the Poynting flux and the kinetic energy of the particles, should lie in the range $10^{-3}\leq \sigma \leq 10^{-2}$  at the termination shock.

Low values are required to avoid the excessive axial compression of high $\sigma$ 1- and 2-D models, that push the pulsar wind termination shock too close to the pulsar \citep{Komissarov.2013, Lyubarsky2012}.
Nevertheless, as recently pointed out by \cite{Porth}, 3D high-$\sigma$ models of PWN have the same morphology of two-dimensional axisymmetric low-$\sigma$ models owing to the presence of kink instabilities \citep{Begelman} that reduce the axial compression and lead to uniform pressure within the Nebula.
In addition, a lateral dependence of the wind magnetization, that increases towards the axis yielding $\sigma \geq 1$ close to the poles, was proposed by recent investigations \citep{Lyubarsky2012, Komissarov.2013}.
In these models, the Nebula and the jet would be therefore injected with a highly magnetized plasma, and they would be regions of strong magnetic dissipation \citep{Komissarov.2013, Porth}.

Unexpectedly, the Crab Nebula produces strong and day long $\gamma$-ray flares \citep{tavani1, abdo, Striani1, Buehler, Striani2}.
The average Nebula spectrum, that shows a cutoff around $100$ MeV, is interpreted as synchrotron emission of electrons and positrons accelerated
at the termination shock.
The $\gamma$-ray flares show, instead, an energy peak at $E_{F}\simeq 400$ MeV.
Magnetic fields of $B\simeq 2$ mG\footnote{The Crab Nebula average magnetic field is B $\simeq 0.2$ mG.} and particles' Lorentz factors $\gamma \simeq 3 \-- 4 \times 10^9$ \citep{Striani2} are  required to match the observable data for the flares.
A very fast and efficient acceleration process is required to explain this very short time-scale and the hard energy peak \citep{tavani1}.

\begin{figure}
 \centering
  \includegraphics*[width=0.5\textwidth]{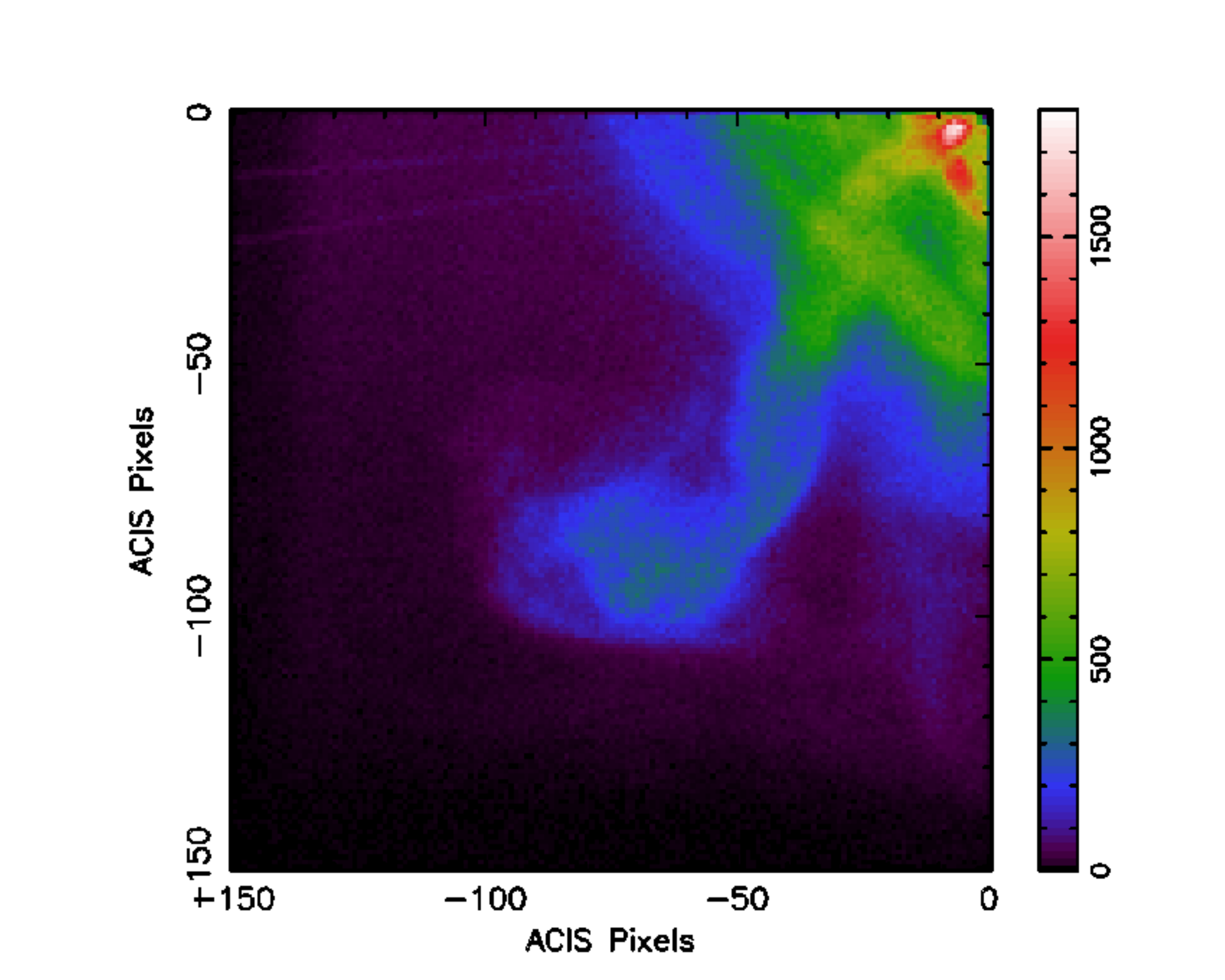}
  \caption{\small Summed image for 7 Chandra ACIS observations of the Crab Nebula's southern jet.
  The data were taken between 2010, September and 2011, April.
  The color bar gives the summed counts per ACIS pixel over a total effective exposure of about 4.2 ksec.
  One ACIS pixel is 0.492 seconds of arc. North is up, East is to the left,
  and the pulsar is at (0, 0) in the figure.
These data are from an exhaustive study of all of the Chandra observations (Weisskopf et al. in preparation).}
  \label{fig:crab_bent_jet}
\end{figure}
The Chandra images of the Crab Nebula show a strongly bent jet \citep{Weisskopf}, see Fig. \ref{fig:crab_bent_jet}.
Interestingly, the comparison of the 2001 and 2010 Chandra images show that
the deflection changed its orientation in this lapse of time \citep{Weisskopf2011}\footnote{A bent jet seems to be a common feature of jets in PWN, as shown, e.g., by the jets of the Vela PWN  \citep{Durant2013}}.
This phenomenon can be the consequence of kink instabilities taking place in the jet.
Previous MHD simulations of the Crab Nebula \citep[see, for instance][]{KL.2003,KL.2004, dZAB.2004,dZVAB.2006} assumed a 2-D axisymmetric model, that does not allow deformations of the jet induced by current-driven (CD) modes with azimuthal wave number $m\neq 0$.

In this paper, for the first time, we present 3-D relativistic MHD simulations of the Crab Nebula jet and investigate for which values of the magnetization $\sigma$ and Lorentz factor $\gamma$ our simulations reproduce the observed jet behavior.
In this context, previous numerical simulations by \citet{MRBFM.2010} have demonstrated that relativistic jets possessing an axial current can be significantly affected by the growth of non-axial symmetric perturbations leading to large deflection off the main longitudinal axis.
Similar results have been confirmed by other investigators, e.g., \cite{MSO.2008} (for Newtonian MHD) and \cite{Porth.2013} (for relativistic MHD) who have studied the growth of current driven modes in magnetocentrifugally driven jets through three dimensional simulations.
Likewise, \cite{Mizuno.2011} have demonstrated the importance of three-dimensional dynamics on the stability of a hot plasma column threaded by a toroidal magnetic field.

Besides, kink instabilities in the jet could trigger magnetic reconnection episodes that could eventually be responsible for the observed $\gamma$-ray flares.
Magnetic reconnection in current sheets created in the torus and/or the jet is, indeed, considered as one of the possible mechanism for the acceleration of particles at PeV energies (e.g., \cite{Cerutti, Uzdensky, Cerutti2013}).

The paper is organized as follows: in section \ref{sec:setup} we describe the initial model setup and its connection to the 2D axisymmetric models of (\citet{dZAB.2004}, dZAB04 henceforth) and describe the relevant simulation parameters.
In section \ref{sec:results} we present and discuss our results while conclusions are drawn in section \ref{sec:summary}.

\section{Model Setup}
\label{sec:setup}
%
%
%
%

\subsection{Equations and Method of Solution}
%

In what follows we consider an ideal relativistic magnetized fluid with rest-mass density $\rho$, bulk velocity $\vec{v}$, magnetic field $\vec{B}$ and thermal (gas) pressure $p$.
Numerical simulations are carried out by solving the equations of special relativistic MHD (RMHD) in conservation form \citep{Anile90}:
\begin{equation}
\label{eq:rmhd}
\begin{array}{lcl}
\DS \pd{(\rho\gamma)}{t}
   + \nabla\cdot\left(\rho\gamma\vec{v}\right) &=& 0 \,,
 \\ \noalign{\medskip}
\DS \pd{\vec{m}}{t}        
  + \nabla\cdot\left[
     w\gamma^2\vec{v}\vec{v} - \frac{\vec{B}\vec{B}}{4\pi}
                             - \frac{\vec{E}\vec{E}}{4\pi}\right]
  + \nabla p_t &=&0  \,,
 \\ \noalign{\medskip}
\DS \pd{\vec{B}}{t} - \nabla\times\left(\vec{v}\times\vec{B}\right) &=& 0  \,,
\\ \noalign{\medskip}
\DS \pd{\cal E}{t}  
  + \nabla\cdot\left(\vec{m}-\rho\gamma\vec{v}\right) &=& 0  \,,
\end{array}
\end{equation}
where $\gamma$ is the Lorentz factor, $\vec{m} = w\gamma^2\vec{v} + \vec{E}\times\vec{B}/(4\pi)$ is the momentum density, $\vec{E}=-\vec{v}\times\vec{B}$ denotes the electric field and $w$ is the gas enthalpy which relates to $\rho$ and $p$ via the ideal gas law:
\begin{equation}\label{eq:gammaEos}
  w = \rho + \frac{\Gamma p}{\Gamma-1} \,.
\end{equation}
We adopt here $\Gamma = 4/3$ appropriate for a hot relativistic plasma.
Total pressure and energy include thermal and magnetic contributions and can be written as
\begin{equation}
 p_t = p + \frac{\vec{B}^2 + \vec{E}^2}{8\pi} \,,\quad
 {\cal E} = w\gamma^2 - p +\frac{\vec{B}^2 + \vec{E}^2}{8\pi} - \rho\gamma\,.
\end{equation}
%
Note that velocities are naturally expressed in units of the speed of light $c=1$.

An additional equation, describing the advection of a passive scalar (or tracer) ${\cal T}(x,y,z,t)$, is included to discriminate between jet material (where ${\cal T}=1$) and the environment (where ${\cal T}=0$):
\begin{equation}\label{eq:tracer}
 \pd{\cal T}{t} + \vec{v}\cdot\nabla{\cal T} = 0\, .
\end{equation}

The system of Equations (\ref{eq:rmhd}) is solved in the laboratory frame using the PLUTO code for astrophysical gasdynamics \citep{Pluto.2007} with linear reconstruction and a second-order TVD Runge-Kutta time stepping with a Courant number of $C_a=0.3$.
PLUTO is a shock-capturing code targeting supersonic fluid in presence of discontinuous waves and largely based on the usage of Riemann solvers.
For the present purpose we employ the HLLD Riemann solver of \citet{MUB.2009} and revert to the HLL solver in presence of strong shocks following the hybrid approach described in the appendix of \cite{Pluto.2012}.
The divergence-free condition of the magnetic field is accurately treated using the constrained transport method.

\subsection{Initial and Boundary Conditions}
%

\begin{figure}
  \includegraphics[width=0.5\textwidth]{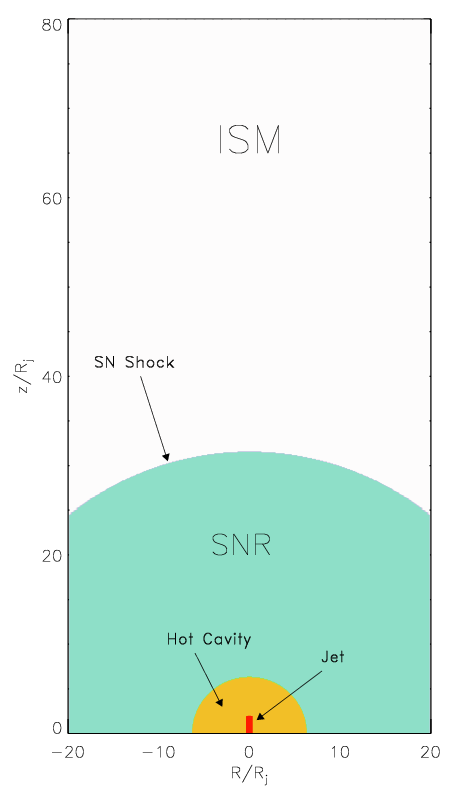}
  \caption{\small Two-dimensional schematic representation of the initial condition. 
  The jet enters from the nozzle at the bottom boundary into a hot cavity region confined by the supernova remnants (SNR) expanding into the outer interstellar medium (ISM).  
  Tick labels are given in units of the jet radius $R_j$.
}
  \label{fig:setup}
\end{figure}

Our initial conditions draw upon the same configuration used by dZAB04 where a freely expanding supernova remnant initially fills the region $0.2 < r/r_{\rm ej} < 1 $ where $r$ is the spherical radius and $r_{\rm ej} = 1 \,{\rm ly}$, see Fig \ref{fig:setup}.
The supernova ejecta is unmagnetized with total mass given by
\begin{equation}
 M_{\rm ej} = \int_0^1 \rho_{\rm ej} 4\pi r^2\, dr  \,,
\end{equation}
and a radially increasing velocity profile $v_r=v_{\rm ej} r/r_{\rm ej}$ where
$v_{\rm ej}$ is fixed by the condition
\begin{equation}
  E_{\rm ej} = \int_0^1\HALF\rho_{\rm ej} v^2 4\pi r^2\, dr \,.
\end{equation}
Here we take $M_{\rm ej} = 3 M_\odot$ and $E_{\rm ej} = 10^{51}\, {\rm erg}$ in accordance with Eq (5) and (6) of \citet{dZAB.2004}.
Further out, for $r/r_{\rm ej} > 1$, the fluid is uniform and static with density and pressure values representative of the interstellar medium (ISM), i.e., $\rho_{\rm ism} = 1\,m_p\,{\rm cm}^{-3}$ and $p=10^{-9}\rho_{\rm ism}c^2$, where $m_p$ is the proton mass.

Differently from dZAB04, however, we do not include the pulsar wind nebula for $r/r_{\rm ej} < 0.2$ but consider, instead, a static hot plasma region where jet acceleration is assumed to take place.
Density and pressure in this region are denoted with $\rho_a$ and $p_a$ while the magnetic field is absent, see also Fig \ref{fig:setup}.

As the acceleration mechanism cannot be consistently described within our model, we assume that the jet has already formed as the result of the magnetic hoop stress and collimation processes taking place around the polar axis.
For this reason, the jet is modeled as a continuous injection of mass, momentum, magnetic field and energy from the lower $z$-boundary inside the circular nozzle $R < R_j$ where $R=\sqrt{x^2+y^2}$ is the cylindrical radius and $R_j=3\cdot 10^{16}\, \rm{cm}$ is the jet radius.
Here inflow values are prescribed in terms of constant density $\rho_j=\rho_a$ and axial velocity
\begin{equation}
 v_z(R) = \sqrt{1 - \frac{1}{\gamma_j^2}} 
\end{equation}
where $\gamma_j$ is the bulk Lorentz factor.

Since present axisymmetric models of PWN adopt a purely toroidal field we initialize the magnetic field to be azimuthal with the following radial profile \citep{Kom.1999, MUB.2009}:
\begin{equation}\label{eq:Bphi}
 B_\phi(R) = \left\{\begin{array}{ll}
\DS B_m \frac{R}{a}  & \quad\mathrm{for}\quad R \le a \,,\\ \noalign{\medskip}
\DS B_m \frac{a}{R}  & \quad\mathrm{for}\quad R > a \,,\\
 \end{array}\right.
\end{equation}
where $a$ encloses a cylinder carrying a constant current and $B_m$ sets the magnetic field strength.
We take $a=R_j/2$ and fix the value of $B_m$ from the jet magnetization parameter $\sigma$:
\begin{equation}
 \sigma = \frac{\av{\vec{B}}^2}{4\pi\rho\gamma_j^2} \,,
\end{equation}
where $\av{\vec{B}}^2 = B_m^2 a^2 (1-4\log a)/2$ is the average magnetic energy inside the beam.

The gas pressure is recovered from the radial momentum balance across the jet which, in absence of rotations, takes the form
\begin{equation}
 \frac{dp}{dR} + \frac{1}{R^2}\frac{d}{dR}
        \left(\frac{R^2}{8\pi}\frac{B_\phi^2}{\gamma^2}\right) = 0 \,,
\end{equation}
where, for simplicity, a constant value of $\gamma=\gamma_j$ is used.
The previous equation has solution
\begin{equation}\label{eq:pressure}
 p(R) = \left\{\begin{array}{ll}
\DS p_a + \frac{B_m^2}{4\pi\gamma_j^2}\left(1 - \frac{R^2}{a^2}\right) & \quad\mathrm{for}\quad R \le a\,,
  \\ \noalign{\medskip}
\DS p_a    & \quad\mathrm{for}\quad R > a\,,
 \end{array}\right.
\end{equation}
where $p_a$ is the ambient pressure determined by flow Mach number $M_s = v_z/c_{s,amb}$.
Note that in this configuration the pressure is maximum on the axis and monotonically decreases until $R=a$ where it matches the ambient pressure.
The maximum value is obtained from Eq. (\ref{eq:pressure}) with $R=0$ and it takes the value $p_j = p(0) = p_a + B^2_m/(4\pi\gamma_j^2)$.
Consequently, highly magnetized jets also possess larger internal energies, see Table \ref{tab:cases}.

As axisymmetric models of PWN predict hot and hollow jets, we prescribe the jet mass density to be $\rho_j = n_jm_p$ with $n_j=10^{-3}\, {\rm cm}^{-3}$ and set the sonic Mach number $M_s=1.7$.
In such a way the initial density contrast between the supernova remnant and the jet is $\approx 10^6$.
Conversely, we leave $\gamma_j$ and $\sigma$ as free parameters and perform computations with two different values of $\gamma_j=2,4$ and three different magnetizations $\sigma = 0.1, 1, 10$ for a total of six cases, as shown in Table \ref{tab:cases}.
This choice of parameters is consistently based on the results obtained from the 2D axisymmetric simulations of dZAB04 (also repeated by our group with good agreement) from which comparable values of density, pressure and magnetic fields could be inferred.

In the injection nozzle, we perturb the transverse velocities by introducing pinching, helical and fluting modes with corresponding wave numbers $m=0,1,2$ \citep{Rossi_etal.2008}:
\begin{equation}\label{eq:jet_perturbation}
  v_R = \frac{A}{24}
   \sum_{m=0}^2\sum_{l=1}^8 \cos\left(m\phi + \omega_lt + b_l\right)
\end{equation}
where $b_l$ are randomly chosen phase shifts while high ($l=1,\dots,4$) and low-frequency ($l=5,\dots,8$)  modes are given by $\omega_l = c_s(1/2,1,2,3)$ and $\omega_l = c_s(0.03, 0.06, 0.12, 0.25)$.
The amplitude of the perturbation is chosen in such a way that the fractional change in the Lorentz factor is $\epsilon = 0.05$:
\begin{equation}
 A = \frac{\sqrt{(1+\epsilon)^2-1}}{\gamma_j(1+\epsilon)}
\end{equation}


The computational domain is defined by the Cartesian box $x,y \in [-L/2,L/2]$ and $z\in[0,L_z]$ with $L=50 R_j$ and $L_z = 80 R_j$ covered by $320\times320\times768$ computational cells.
The grid resolution is uniform in the $z$ direction and inside the region $|x|, |y| < 10$ where $192\times 192$ zones are used.
The grid spacing increases geometrically outside this region up to the lateral sides of the domain.

We employ outflow (i.e. zero-gradient) boundary conditions on the $x$ and $y$ sides of the computational box as well as on the top $z$ boundary. 
In the ghost zones at the bottom $z$ boundary and outside the injection nozzle, we
set $v_z$, $B_x$ and $B_y$ to be anti-symmetric with respect to the $z=0$ plane while the remaining quantities are symmetrized.
Fluid variables inside and outside the nozzle are then smoothly joined using a profile function:
\begin{equation}
  q = q_j + \frac{q_e - q_j}{\cosh\left(R/R_j\right)^n}\,,
\end{equation}
where $q_j$ is a fluid variable value inside the nozzle, $q_e$ is the symmetric (or anti-symmetric) value with respect to the $z=0$ plane.
Finally, we use $n=8$ for density, velocity, pressure and vertical component of the field while $n=6$ is used for $B_x$ and $B_y$.

\begin{table}
 \centering
 \caption{\small Simulation cases describing our initial jet configuration.
         While $\gamma_j$ and $\sigma$ are the primary parameters used in our model we also give the derived values of magnetic field strength ($B_m$, 4-th column), ambient pressure ($p_a$ 5-th column), on-axis pressure ($p_j$, 6-th column) and the ratio between average thermal and magnetic pressures (plasma $\av{\beta}$ , last column).}
  \begin{tabular}[\textwidth]{cc|ccccc}
 \hline
 Case & $\gamma_j$ & $\sigma$ & $B_m/\sqrt{4\pi\rho_jc^2}$  & $p_a/\rho_jc^2$ & $p_j/\rho_jc^2$ & $\av{\beta}$ \\
 \hline
 A1   &     2      &    0.1   & $0.92$ & $0.88$ & $1.09$ & $4.53$
\\  \noalign{\smallskip}
 A2   &     2      &      1   & $2.91$ & $0.88$ & $3.00$ & $0.57$
\\  \noalign{\smallskip}
 A3   &     2      &    10    & $9.21$ & $0.88$ & $22.1$ & $0.18$
\\  \noalign{\smallskip}
 B1   &     4      &    0.1   & $1.84$ & $9.07$ & $9.28$ & $11.37$
\\  \noalign{\smallskip}
 B2   &     4      &      1   & $5.83$ & $9.07$ & $11.2$ & $1.17$
\\  \noalign{\smallskip}
 B3   &     4      &    10    & $18.4$ & $9.07$ & $30.3$ & $0.15$
 \\ \hline
\end{tabular}
\label{tab:cases}
\end{table}

\section{Results}
\label{sec:results}
%
%
%

\begin{figure*}
  \includegraphics*[width=0.33\textwidth]{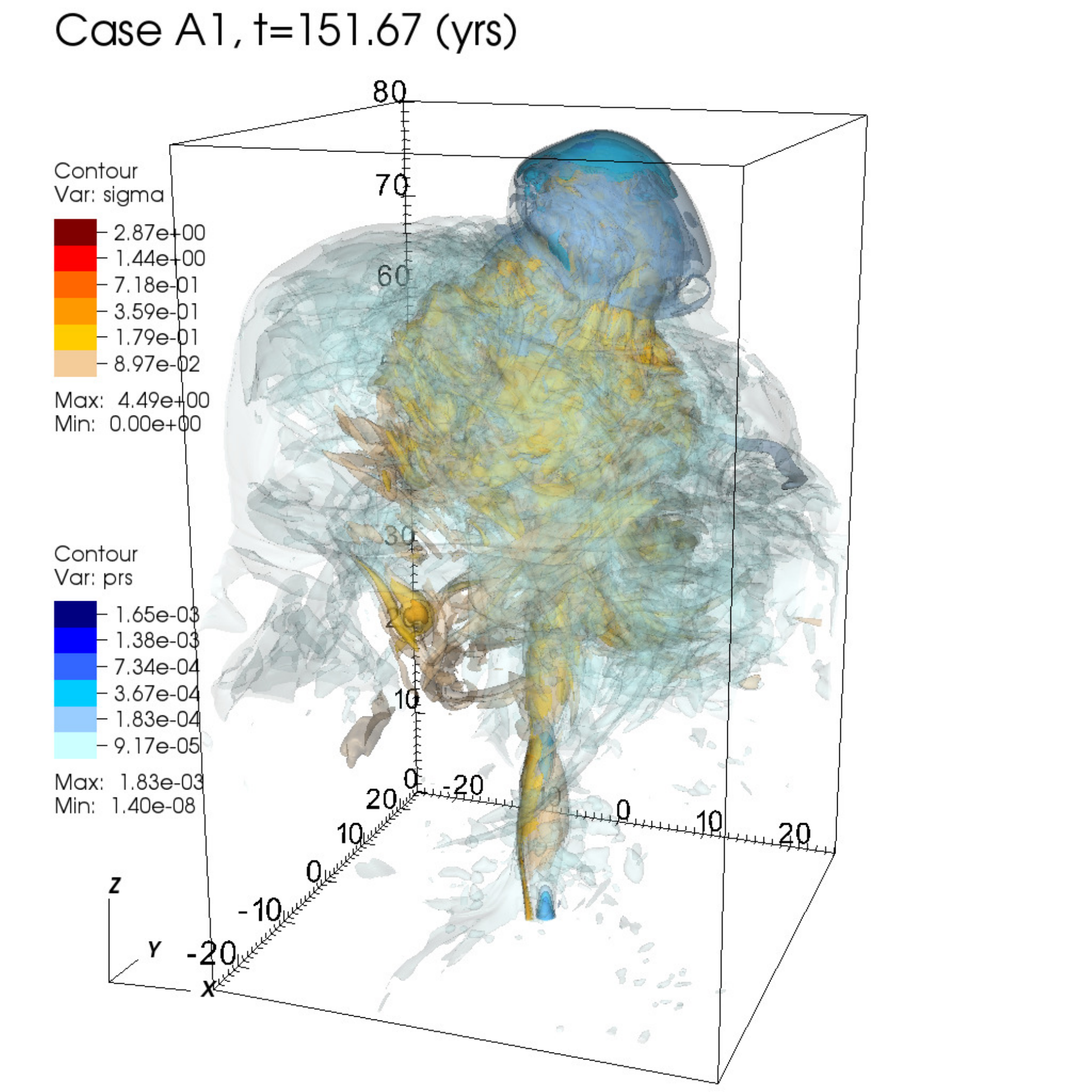}%
  \includegraphics*[width=0.33\textwidth]{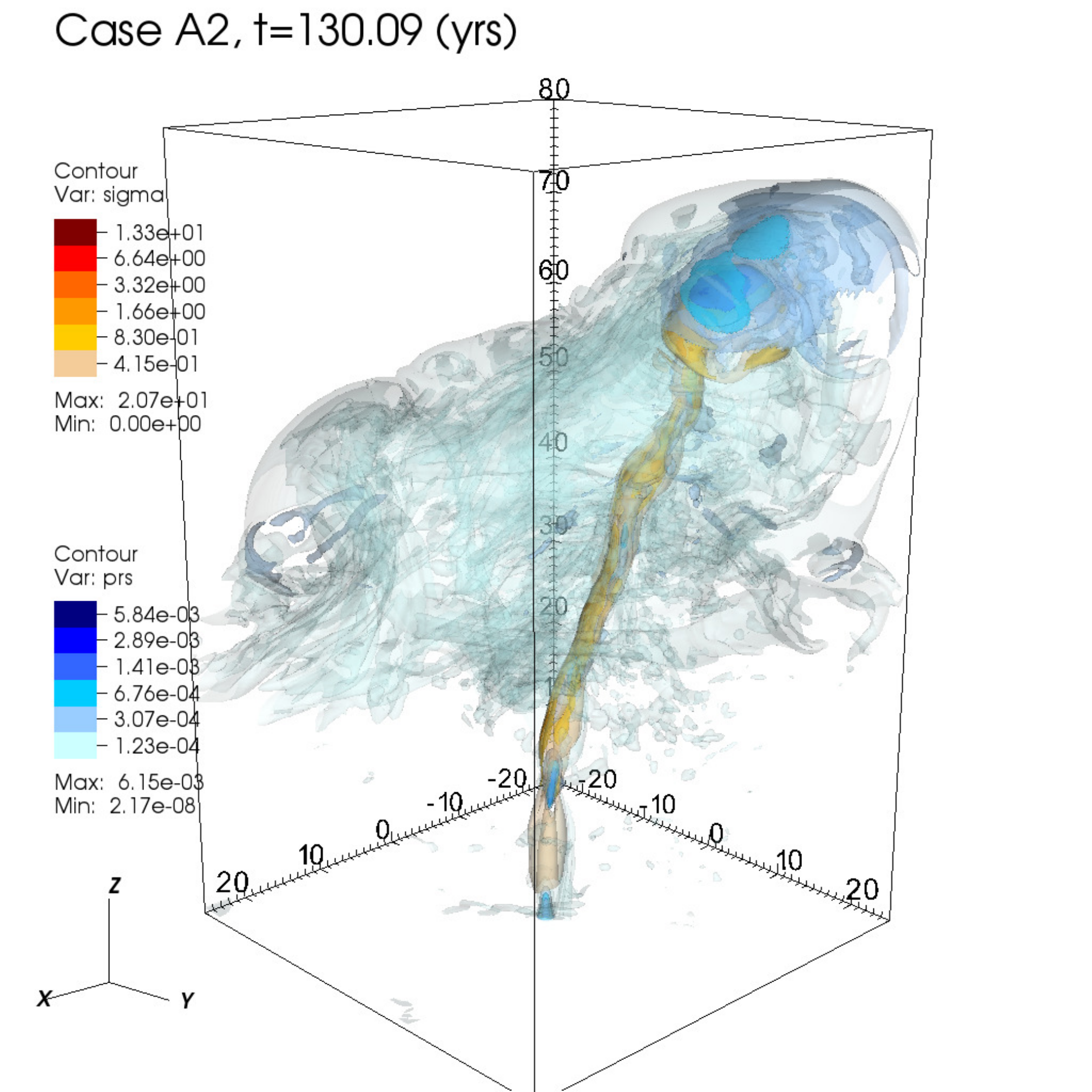}%
  \includegraphics*[width=0.33\textwidth]{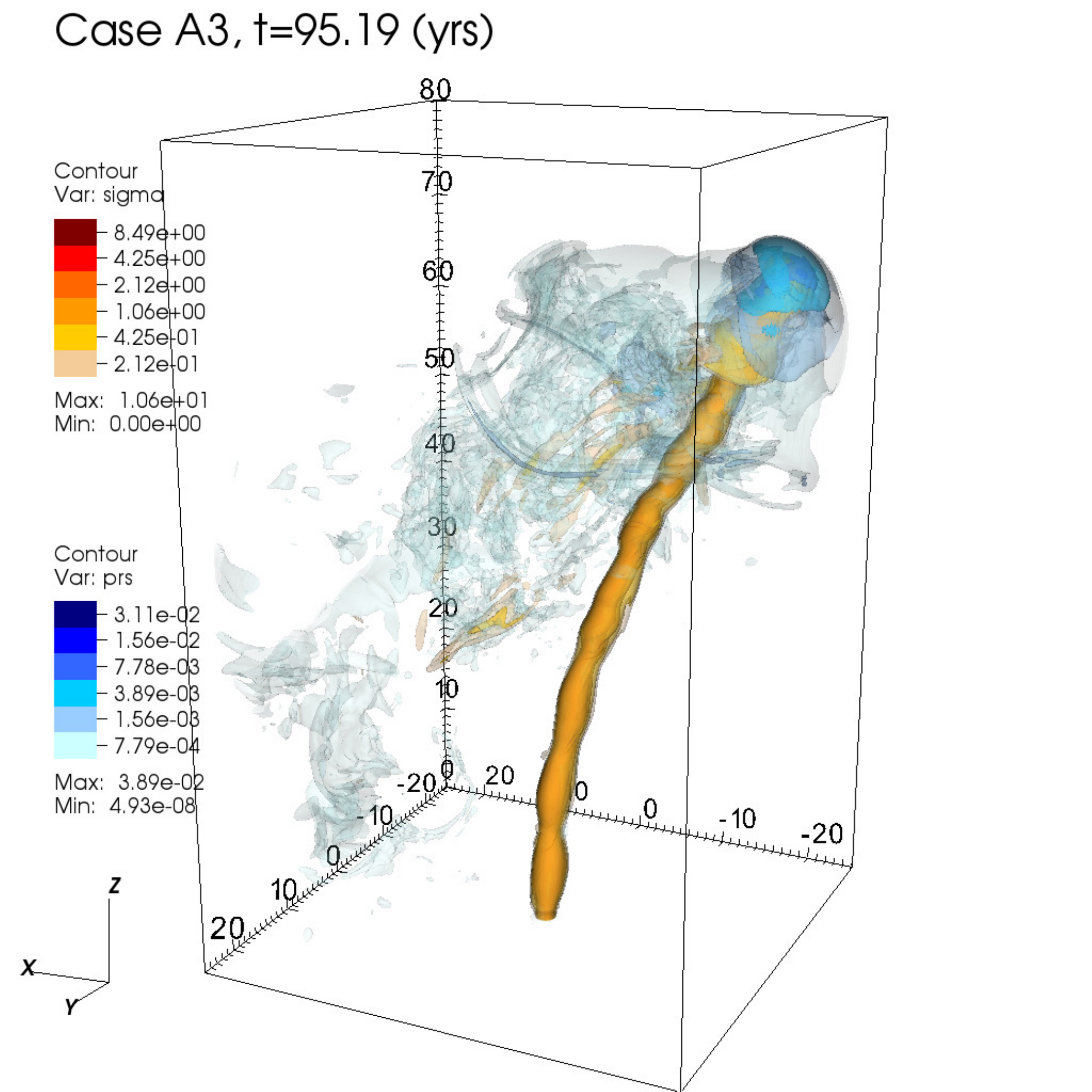}%
  \vspace{1cm}
  \includegraphics*[width=0.33\textwidth]{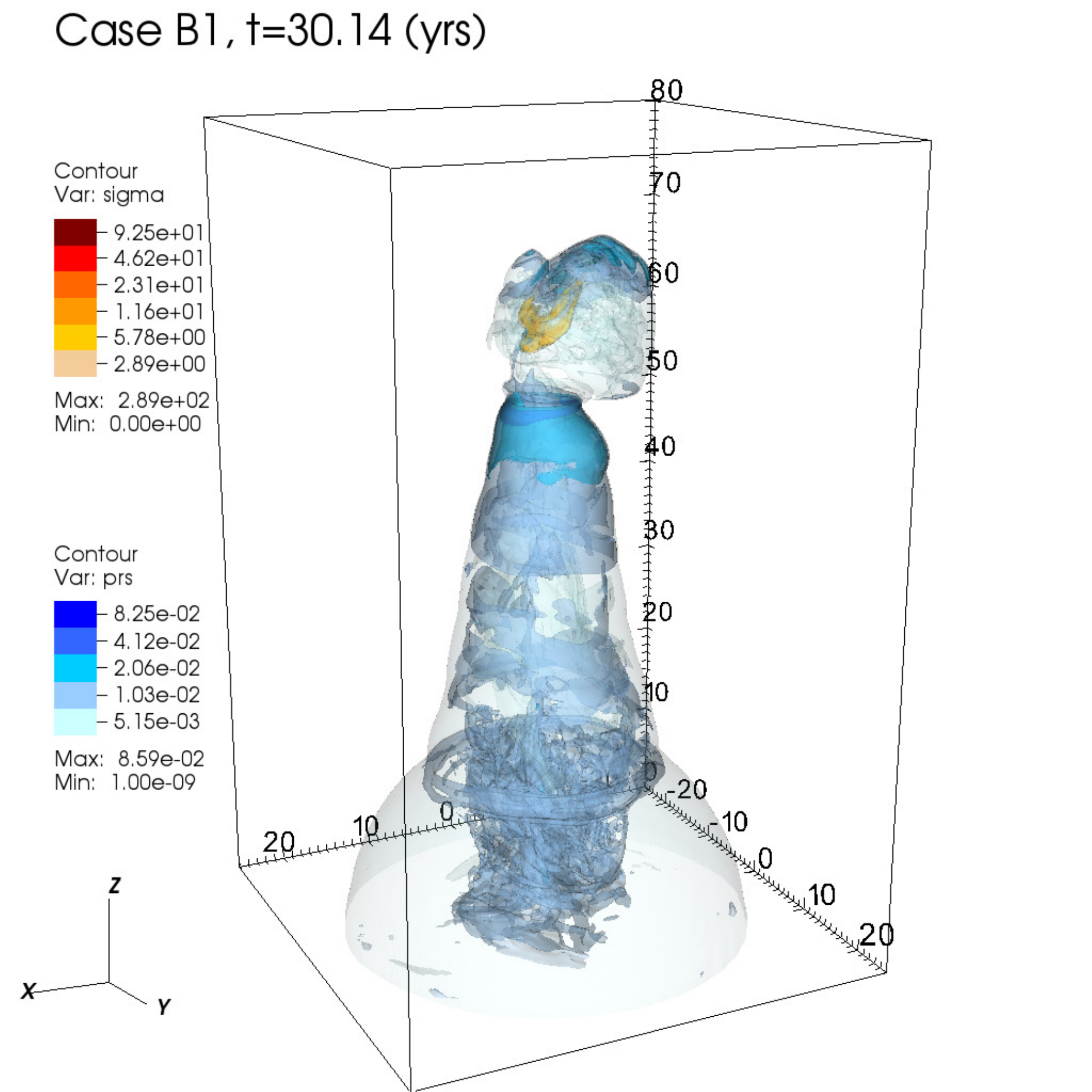}%
  \includegraphics*[width=0.33\textwidth]{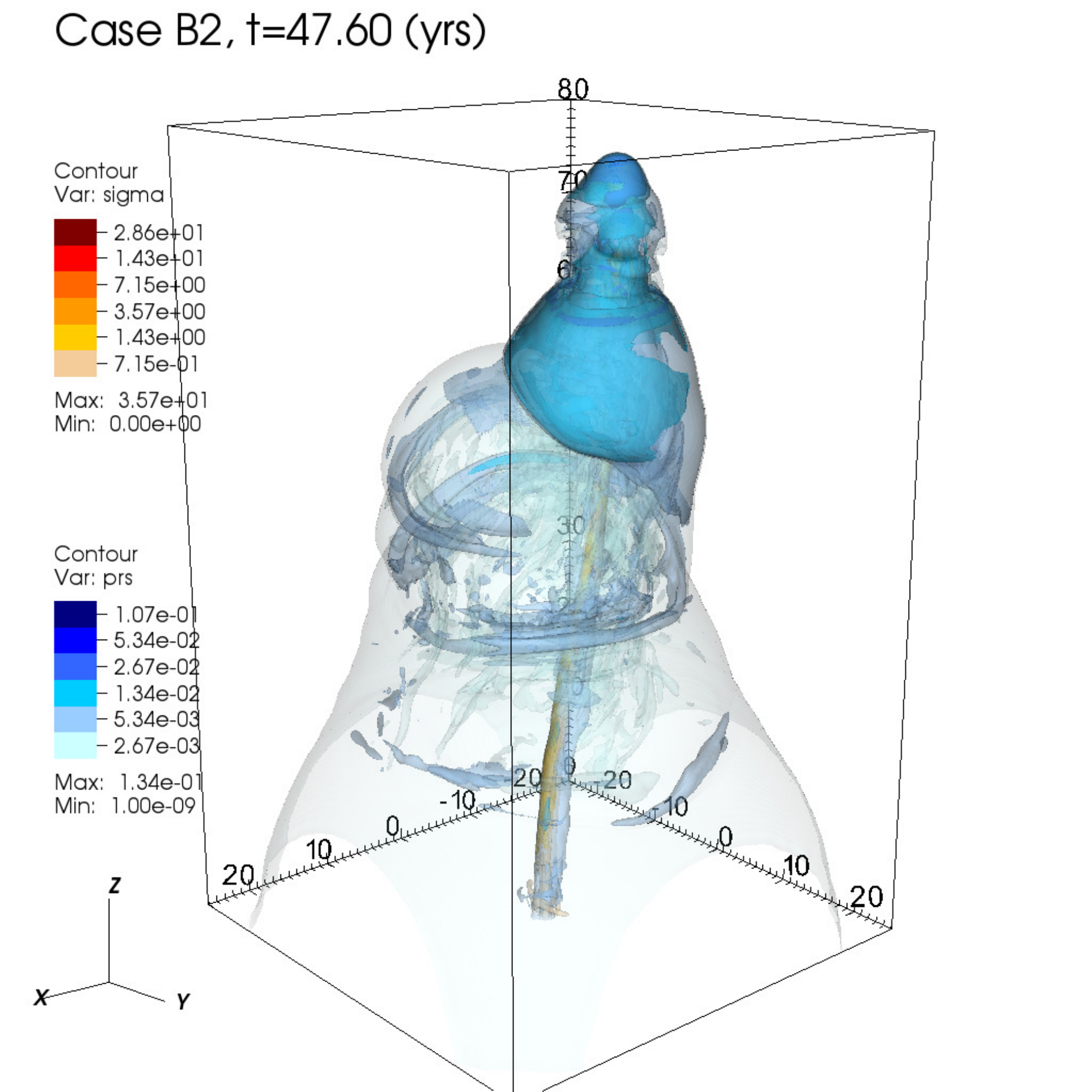}%
  \includegraphics*[width=0.33\textwidth]{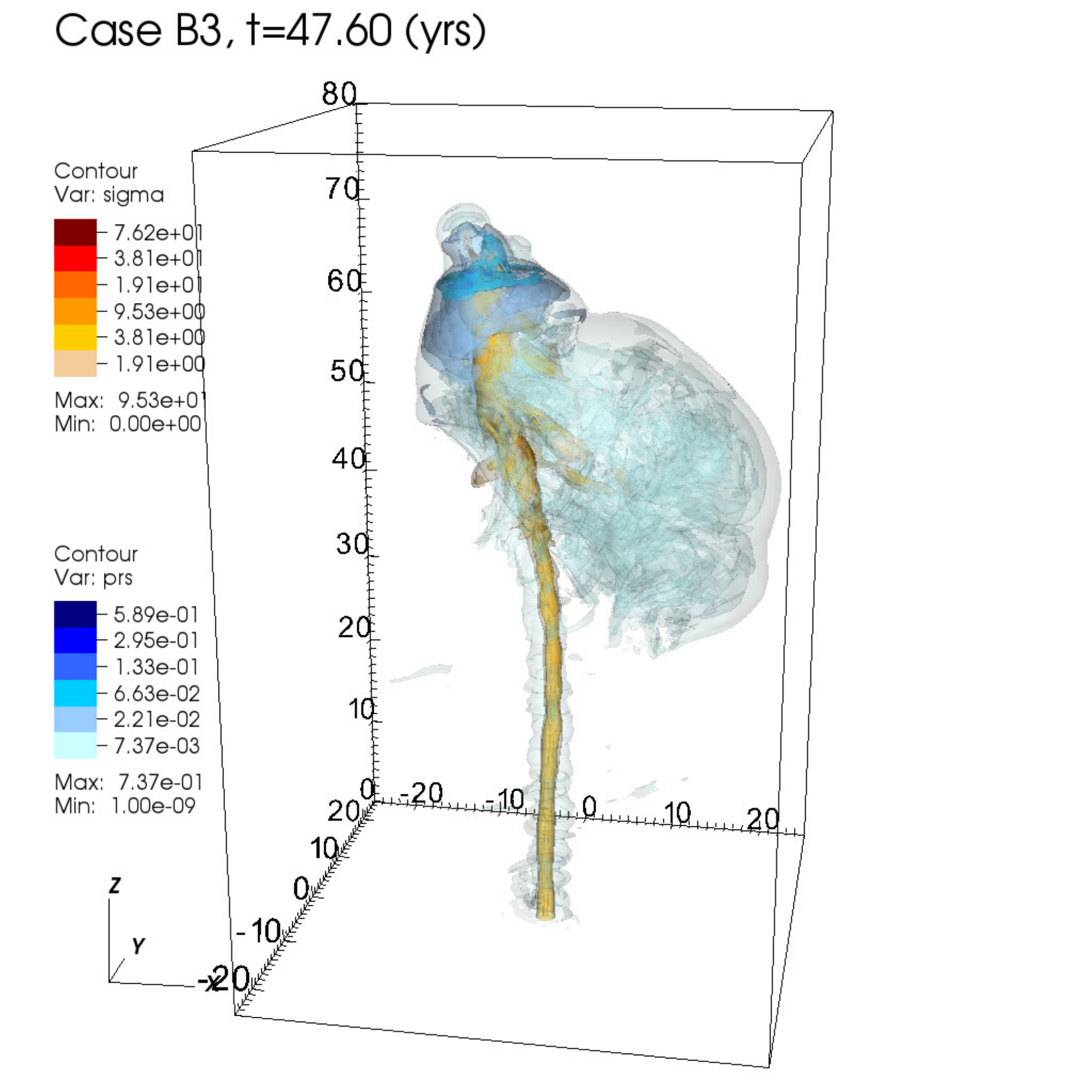}
  \caption{\small Three-dimensional contour surfaces of the gas pressure (blue) and $\sigma$ parameter (orange) for the six jet configurations.
  Tick labels are given in units of the jet radius, $R_j = 3\cdot 10^{16}$ cm.  
  Low-speed jets with $\gamma_j=2$ (cases A1, A2 and A3) are shown, from left to right, in the top panel, while high-speed jets with $\gamma_j=4$ (cases B1, B2 and B3) are shown in the bottom panel.
  Snapshots are taken at different times when the jet has approximately reached the end of the computational domain.}
  \label{fig:final_snapshots}
\end{figure*}

The six simulation cases introduced in Table \ref{tab:cases} show remarkable  differences in several aspects such as the propagation velocity, large-scale morphology, interaction and mixing with the environment.
These will be discussed in the following.

As the computations produced a significant amount of data ($\approx 8 $ TB) an efficient post-processing analysis is crucial in order to reduce and extract relevant quantitative results.
Our experience has shown that several morphological and dynamical aspects can be readily interpreted by means of horizontally-averaged quantities defined as
\begin{equation}\label{eq:average}
  \av{Q}(t,z)\equiv \left<Q,\chi\right> =\frac{\DS\int Q(t,\vec{x})\chi \,dx\,dy}
       {\DS\int\chi \, dx\,dy} \,,
\end{equation}
where integration is performed over horizontal planes at constant-$z$,
$Q(\vec{x},t)$ can be any flow quantity, $\vec{x}=(x,y,z)$ is the position vector and $\chi$ is a weight (or filter) function used to include or exclude certain regions of the flow according to specific criteria.
We select, for instance, computational zones containing more than $50 \%$ of the jet material and moving at least at $25\%$ of the speed of light using
\begin{equation}\label{eq:jet_chi}
 \chi_j = \left\{\begin{array}{ll}
  {\cal T} & \qquad \mathrm{for}\quad {\cal T} > 0.5 \quad \mathrm{and}
 \quad |\vec{v}| > 0.25 \,,
  \\ \noalign{\medskip}
  0  & \qquad \mathrm{otherwise}\,,
\end{array}\right.
\end{equation}
where ${\cal T}$ is the passive scalar obeying equation (\ref{eq:tracer}).

\subsection{Overall Features.}
%

During the very early phases of evolution, the jet propagates almost undisturbed until it impacts the backward dense layers of the supernova remnant.
From this time on (typically $\sim 1 $ year), we observe a drastic deceleration as the jet pushes against the much heavier material of the remnant.

After a few tens of years, the typical structure consists of a large over-pressurized turbulent cocoon enclosing a collimated magnetized central spine moving at mildly relativistic velocities.
This is illustrated in Fig. \ref{fig:final_snapshots} where a volume rendering of thermal pressure and magnetization parameter $\sigma$ is shown for each of the six cases at the end of the simulation\footnote{A collection of movies for the simulation cases presented here can be found at http://plutocode.ph.unito.it/CrabJet/.}. 

The cocoon appears to be weakly magnetized and the field remains mainly concentrated in the beam (a similar structure is also observed in the 2D simulations of dZBL04) preserving the initial toroidal structure.
Along the jet spine the flow experiences a series of acceleration and deceleration phases owing to the presence of conical recollimation shock waves (or working surfaces) corresponding to regions of jet pinching.

The large-scale morphology is characterized by elongated curved structures which, depending on the case, may considerably departs from axial symmetry.
The amount of bending and twisting varies according to the combination the flow Lorentz factor and magnetization ($\sigma$) and it is described in more detail in Section \ref{sec:deflection}.

\subsubsection{Jet Position.}
%

\begin{figure}
 \begin{center}
  \includegraphics[width=0.45\textwidth]{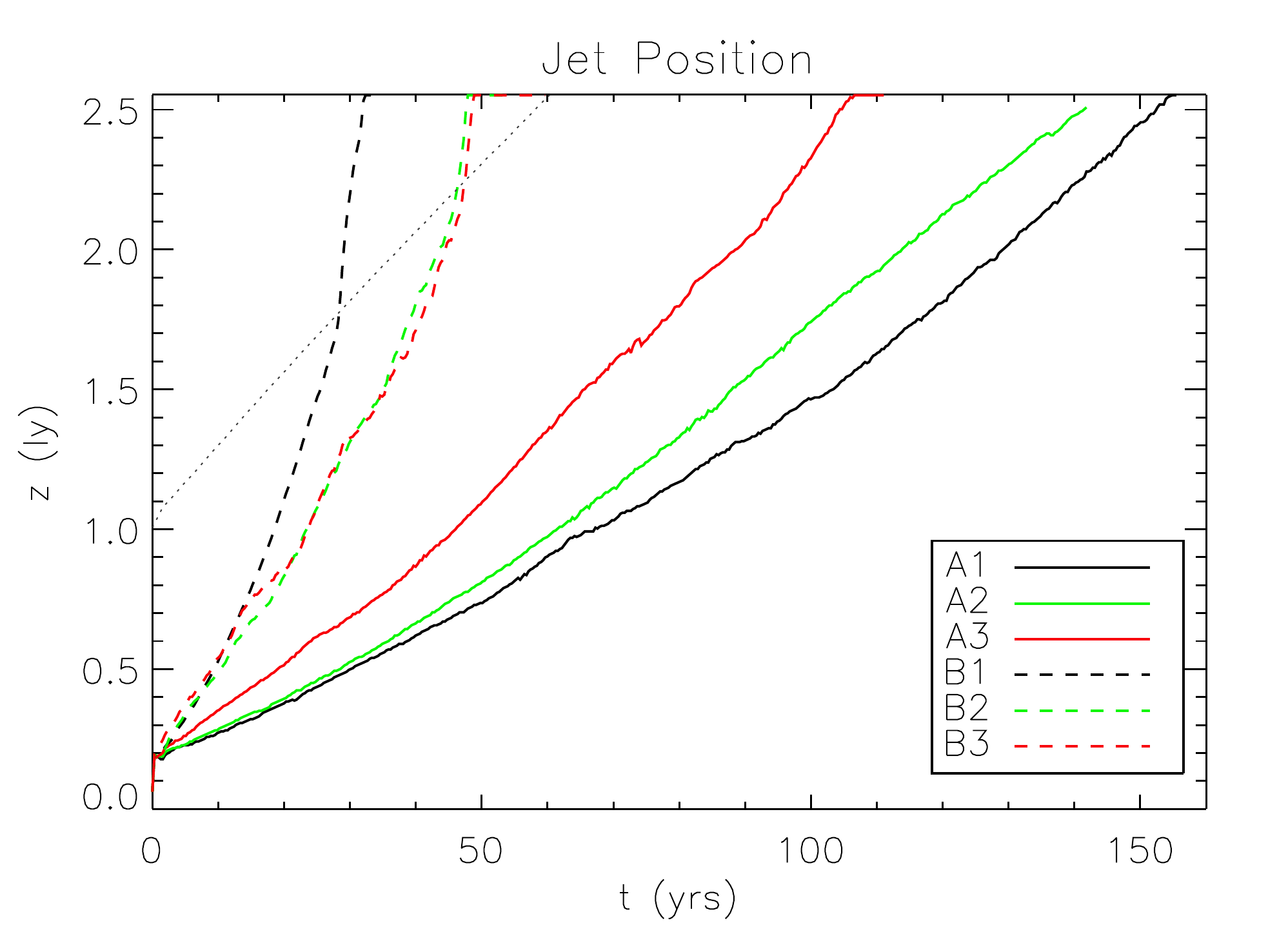}
  \caption{\small Jet head position as function of time for the six simulation cases A1, A2, A3 (black, green and red solid lines) and B1, B2, B3 (corresponding dashed lines).
  The thin dotted line on the left represents the position of the outer supernova shock.}
  \label{fig:jetpos}
 \end{center}
\end{figure}

Fig. \ref{fig:jetpos} shows the jet head position as a function of time, measured as the maximum height $z(t)$ at which jet material has propagated.

Low-speed jets advance slowly ($0.016 \lesssim v_{\rm head}/c \lesssim 0.023$) owing to the large density contrast and evolve entirely within the remnant confined by the outer SN shock (see Fig \ref{fig:setup}).
For increasing magnetization the propagation is driven by the additional magnetic pressure support while the mechanism of instability tends to saturate.

Conversely, jets with larger $\gamma_j$ (cases B1, B2 and B3) advance more rapidly ($0.05\lesssim v_{\rm head}/c \lesssim 0.078$) and cross the outer SN shock (dotted line) at earlier times ($t < 50$ years) where they suddenly accelerate because of the reduced density contrast.
In particular, owing to its low magnetization and relatively large kinetic energy, the B1 jet is very little affected by the growth of CD modes and its trajectory remains essentially parallel to the axis.
With increasing magnetization, the jets in case B2 and B3 are slowed down by appreciable bendings of the flow direction and show comparable propagation speeds.

\subsection{Jet Internal Structure.}
\label{sec:internal_struct}
%

\begin{figure}
 \centering
  \includegraphics*[width=0.45\textwidth]{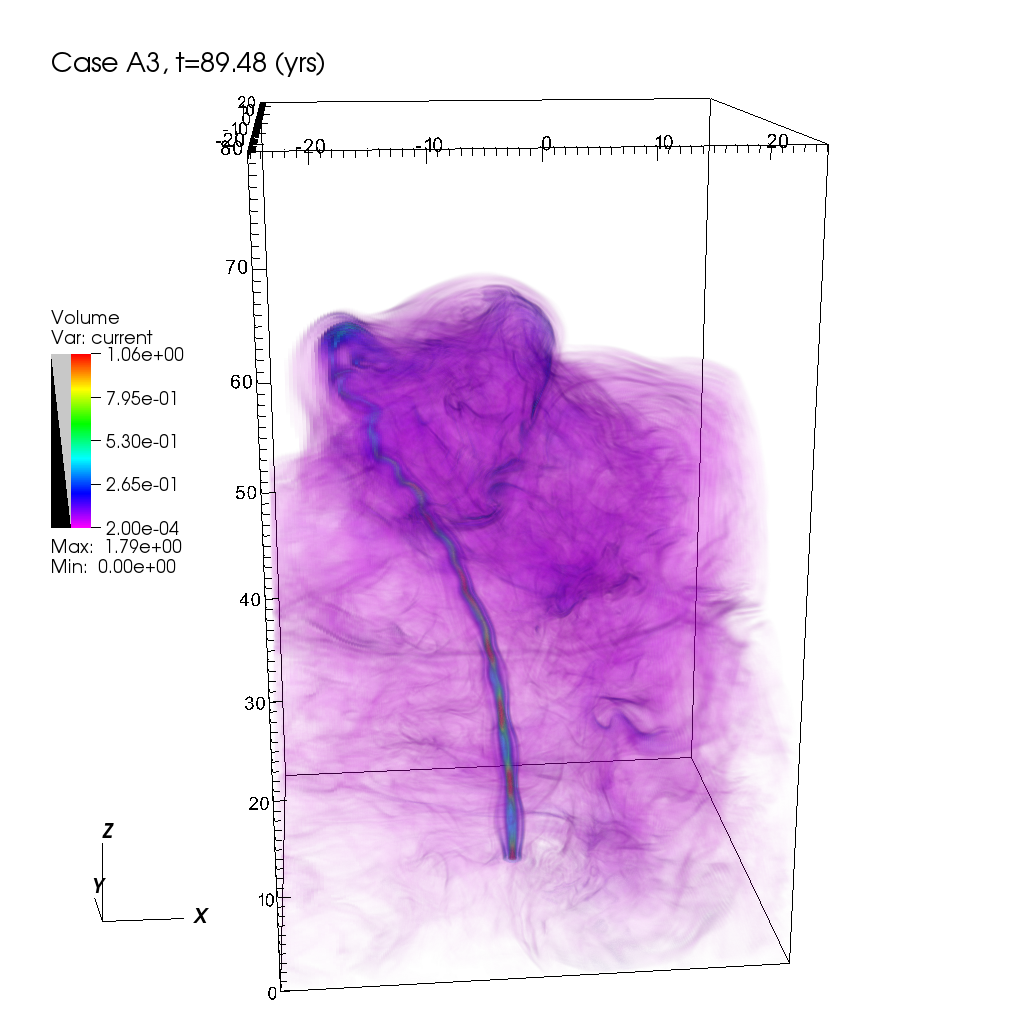}
  \caption{\small Volume rendering of the magnitude of the current density $\vec{J} = \nabla\times\vec{B}/(4\pi)$ for the A3 jet, at $t=89.48\,{\rm yrs}$ showing the formation of helical structures in the front-end regions. }
  \label{fig:A3-helix}
\end{figure}

The jet structure does not remain homogeneous during its propagation but, rather, shows substantial variations of several fluid quantities all along its length.
Broadly speaking, we are able to identify two regions with different properties.
In the back-end region, close to the injection nozzle, the jet has reached a quasi-stationary structure with a number of standing conical shocks.
In the front-end region, the dynamics is characterized by a rapid variability and strong interaction between the jet and the remnant.
In these regions and for large magnetizations, the beam takes the shape of a twisted helical structure, see Fig. \ref{fig:A3-helix} showing a three-dimensional view of the current density for the A3 jet after $\sim 89.5$ years.

\subsubsection{Pinching.}
%

\begin{figure}
  \includegraphics[width=0.45\textwidth]{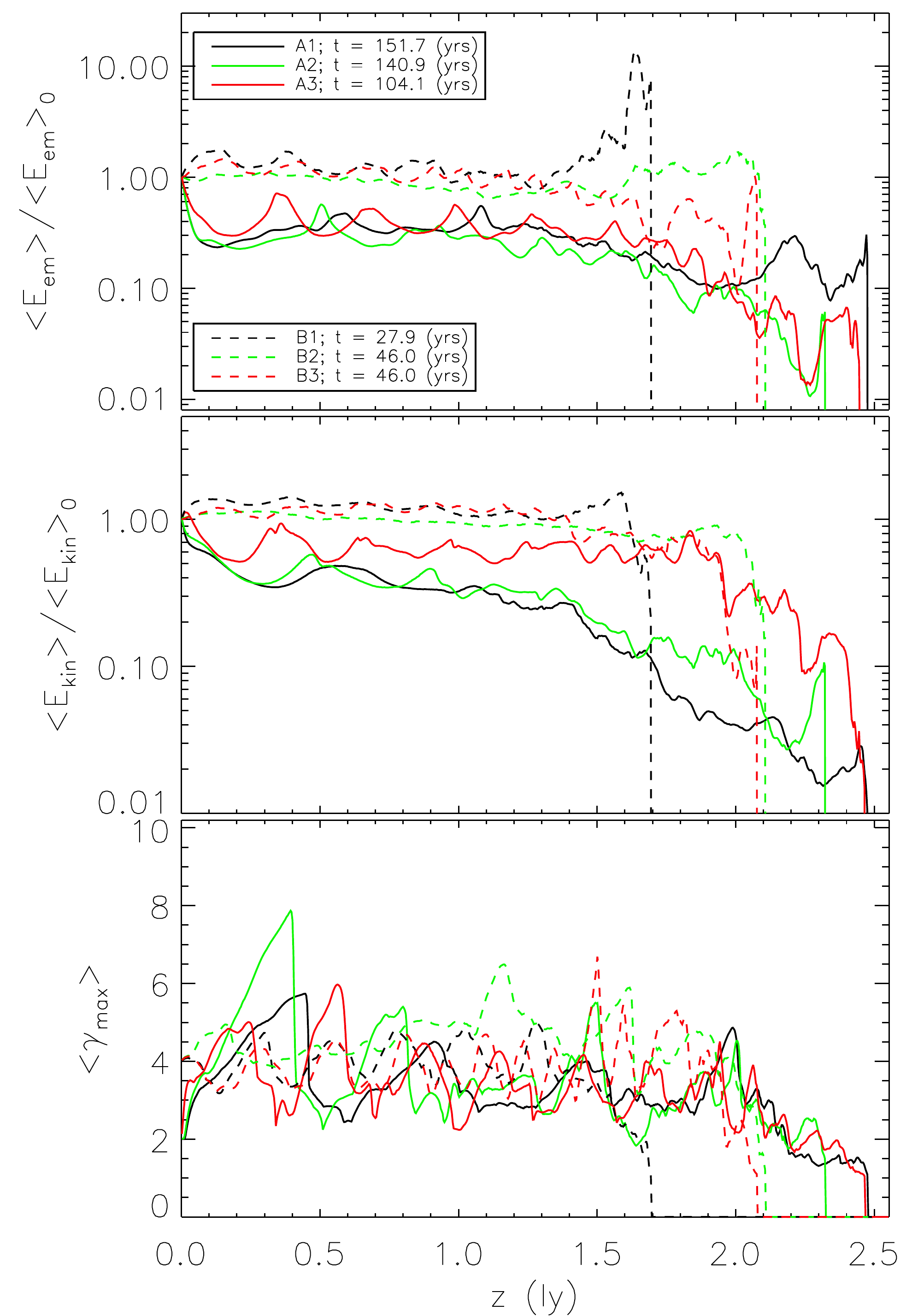}
  \caption{\small Averages profiles of the electromagnetic (top) and kinetic (middle) energies normalized to their initial value at $z=0$ as functions of the vertical distance $z$ at different times (reported in the legend) for the six simulation cases.
  The bottom panel shows the maximum Lorentz factor. 
  Regions of strong compression are evident by the quasi-periodic oscillations.
  The Lorentz factor grows immediately upstream of the shocked flow where magnetic and kinetic energies are smaller and drops discontinuously in the post-shock regions.
}
  \label{fig:Emkl_profiles}
\end{figure}
A common feature that can be identified all along the jet is the presence of pinching regions corresponding to the formation of magnetized shock waves.
These can be distinguished, for instance, by looking at the horizontally-averaged electromagnetic and matter kinetic energies
\begin{eqnarray}\label{eq:Eem+Ekin}
 \av{E}_{\rm em}(t,z) &=& \left<\frac{\vec{B}^2+\vec{E}^2}{8\pi},\chi_j\right>
  \\ \noalign{\medskip}
 \av{E}_{\rm kin}(t,z) &=& \left<\rho\gamma(\gamma-1),\chi_j\right>\,,
\end{eqnarray}
where the weight function $\chi_j$ selects only material that is mainly composed by jet particles, see Eq. (\ref{eq:jet_chi}).

In Fig. \ref{fig:Emkl_profiles} we plot $\av{E}_{\rm em}$, $\av{E}_{\rm kin}$ and the maximum Lorentz factor $\gamma_{\max}$ (taken on $xy$ planes) as functions of $z$ just before the jet has exited the computational domain or encountered the outer supernova shock.
Average magnetic and kinetic energies exhibit quasi-periodic oscillations along the beam due to jet pinching with the corresponding formation of internal shocks with large compression factors.
These cycles are more evident in the slowly-moving jets that reveal shocks with larger strengths.
Here the frequency of oscillations increases with $\sigma$ and magnetic fields tend to dissipate more rapidly.

Indeed, as discussed in \cite{MRBFM.2010}, the presence of a dominant azimuthal magnetic field component prevents the inner jet core from interacting with the surrounding thus substantially reducing the loss and transfer of momentum.
The net effect of this shielding mechanism is to sustain the kinetic energy at the expenses of magnetic energy thus leading to a significant decrease of $\sigma$ along the beam.
\begin{figure}
  \includegraphics[width=0.5\textwidth]{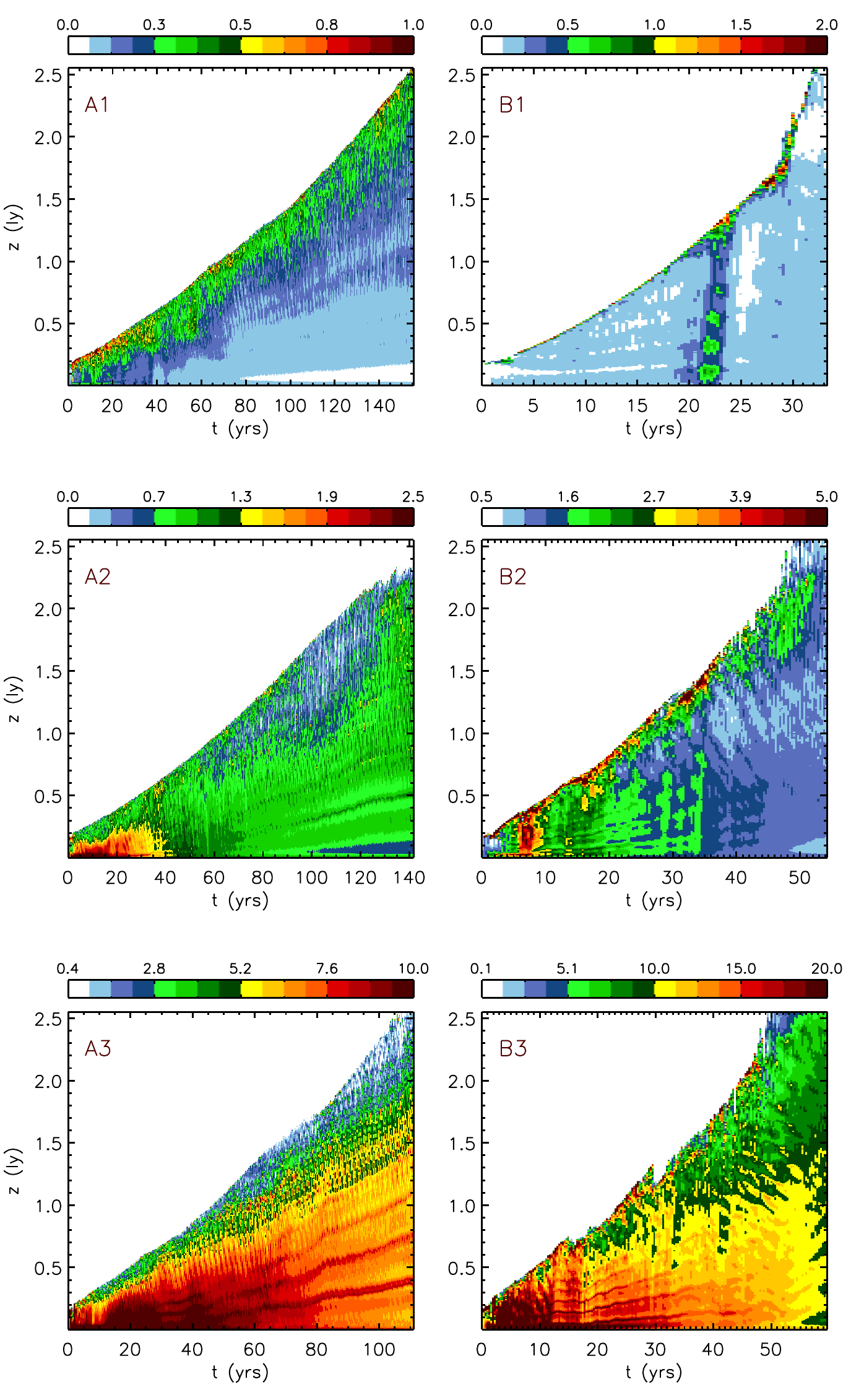}
  \caption{\small Two-dimensional color distribution map of the plane-averaged $\sigma$ parameter normalized to the initial injection value as function of time (abscissa, in years) and vertical height (ordinate, in light-years) for the six  cases.
  Note the substantial decrease of $\sigma$ in the more magnetized cases (A3 and B3).}
  \label{fig:sigma}
\end{figure}
This is best illustrated in Fig. \ref{fig:sigma}, where we show a 2D color distribution map of the horizontally-averaged $\sigma$ parameter normalized to its initial value.

\subsubsection{Fragmentation}
%
%

\begin{figure*}
 \centering
  \includegraphics*[width=0.32\textwidth]{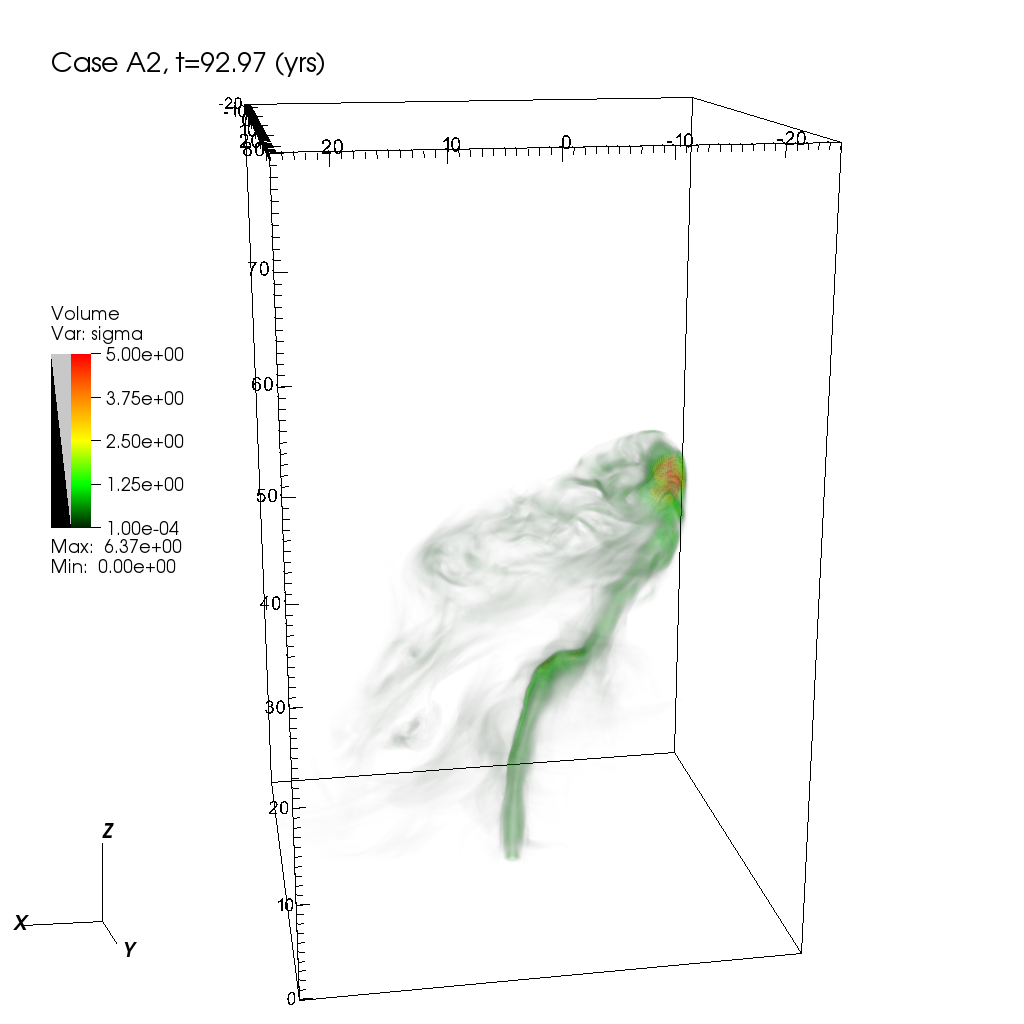}%
  \includegraphics*[width=0.32\textwidth]{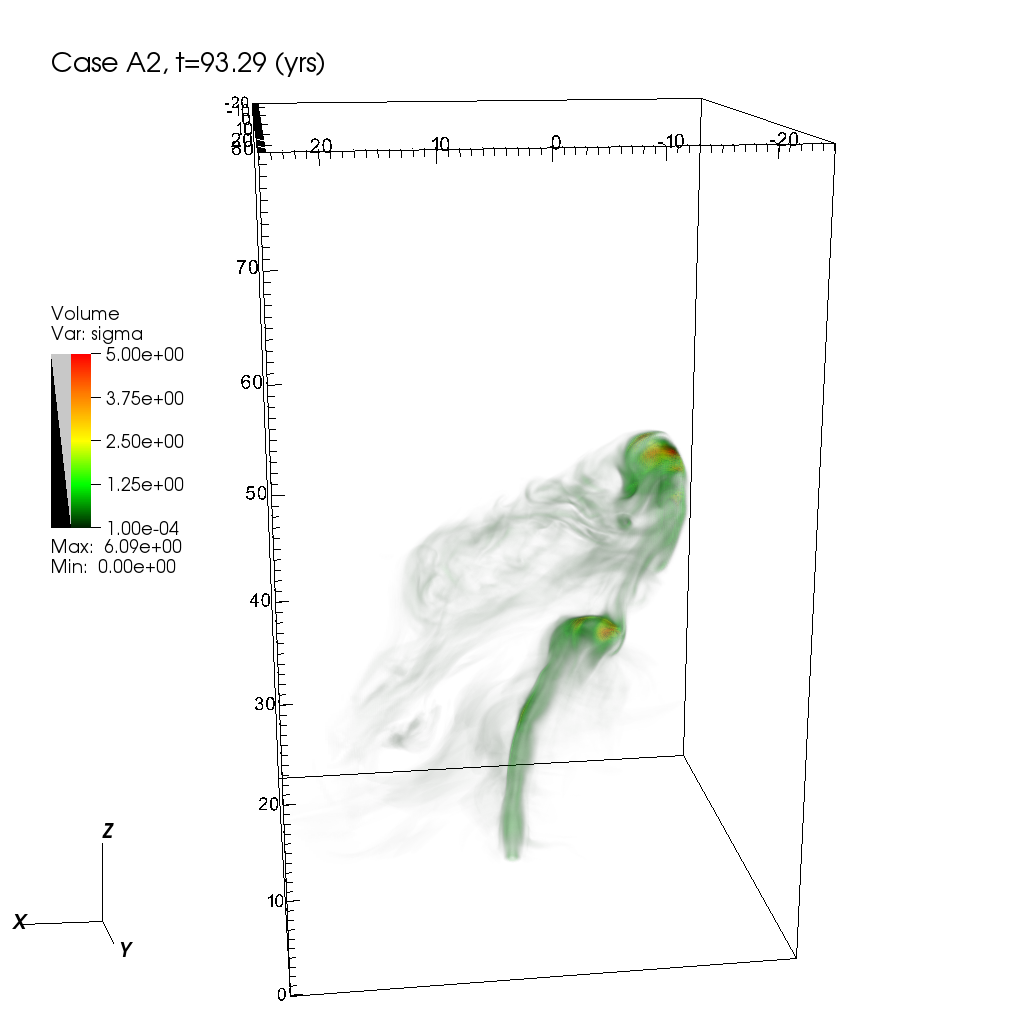}%
  \includegraphics*[width=0.32\textwidth]{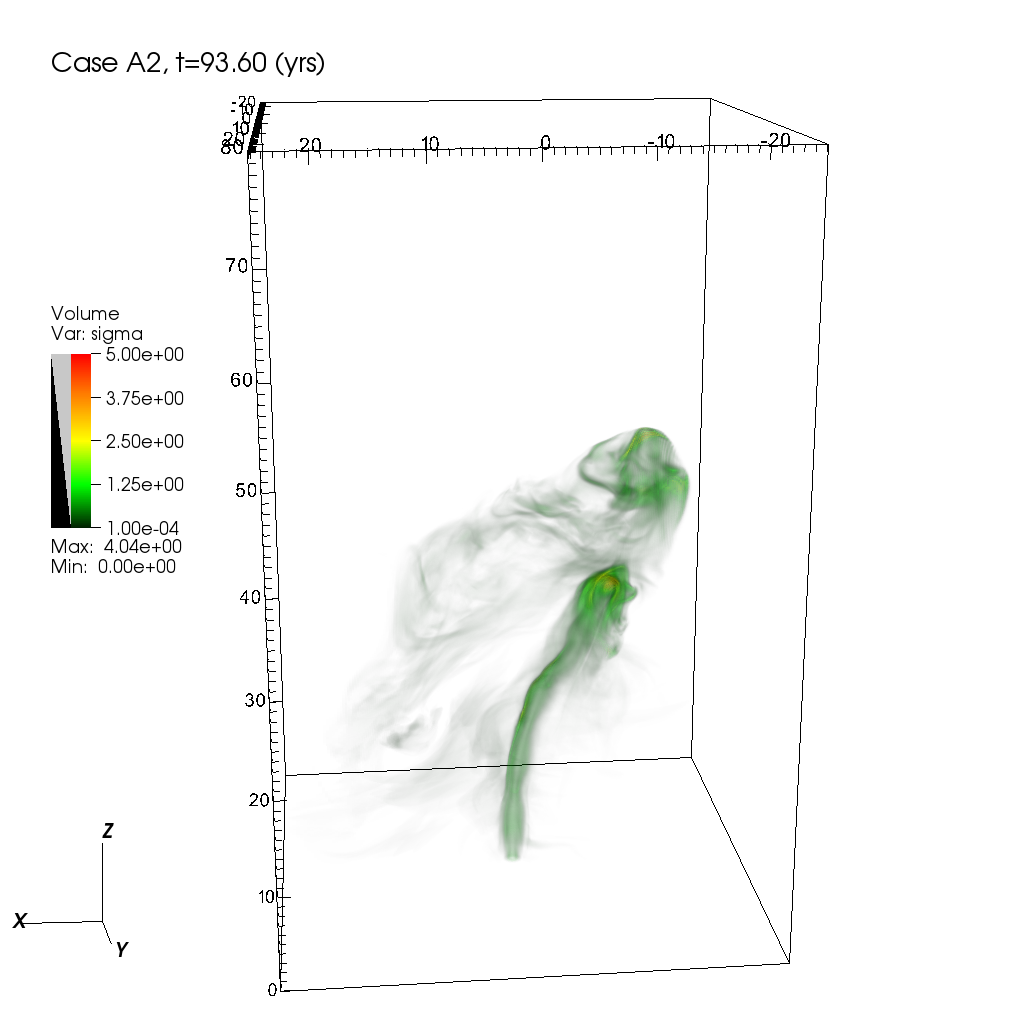}
  \includegraphics*[width=0.32\textwidth]{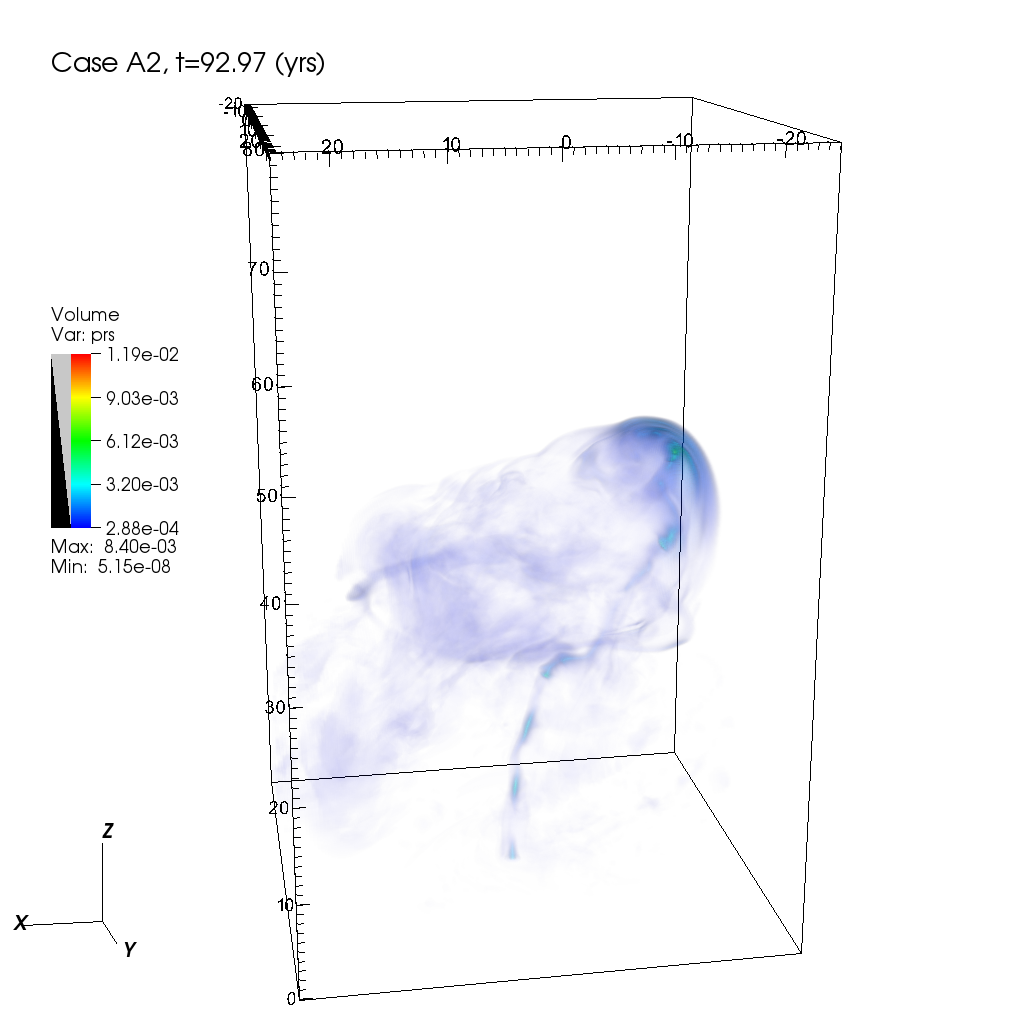}%
  \includegraphics*[width=0.32\textwidth]{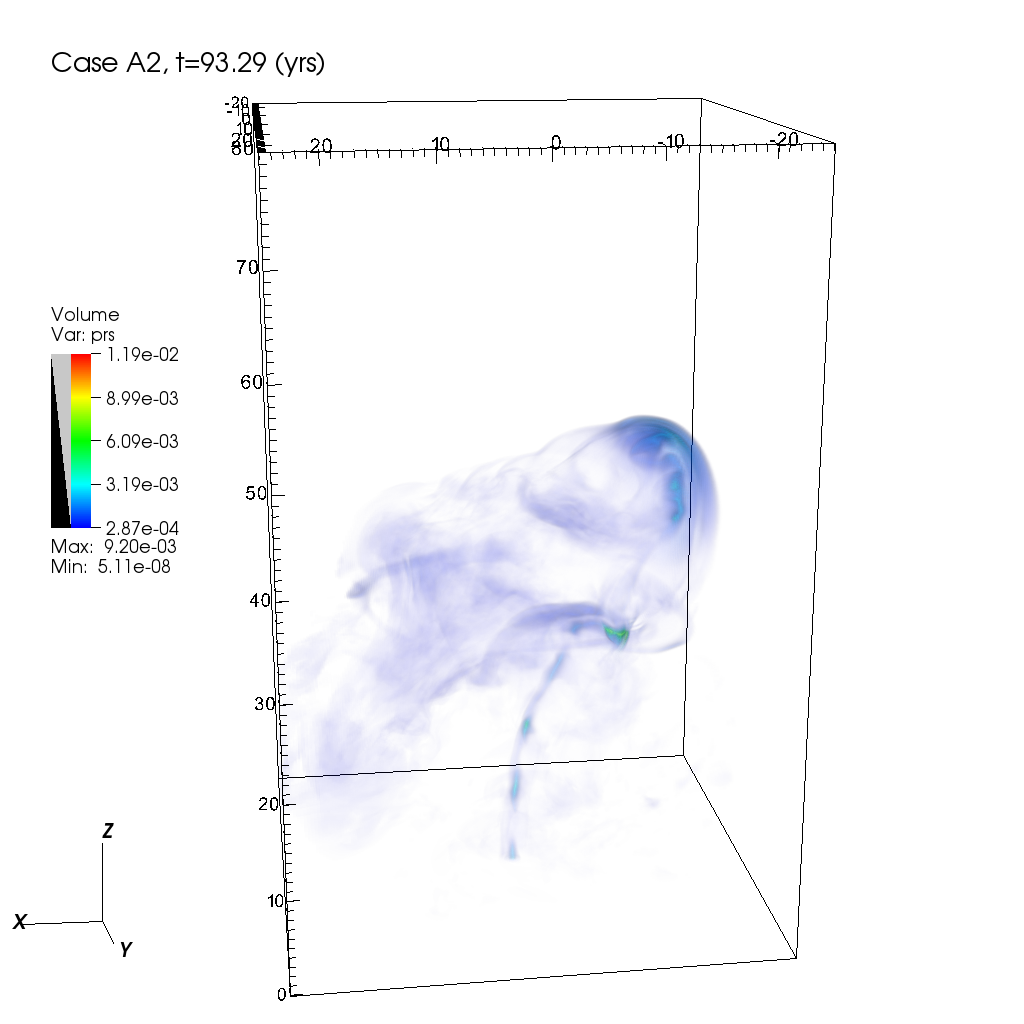}%
  \includegraphics*[width=0.32\textwidth]{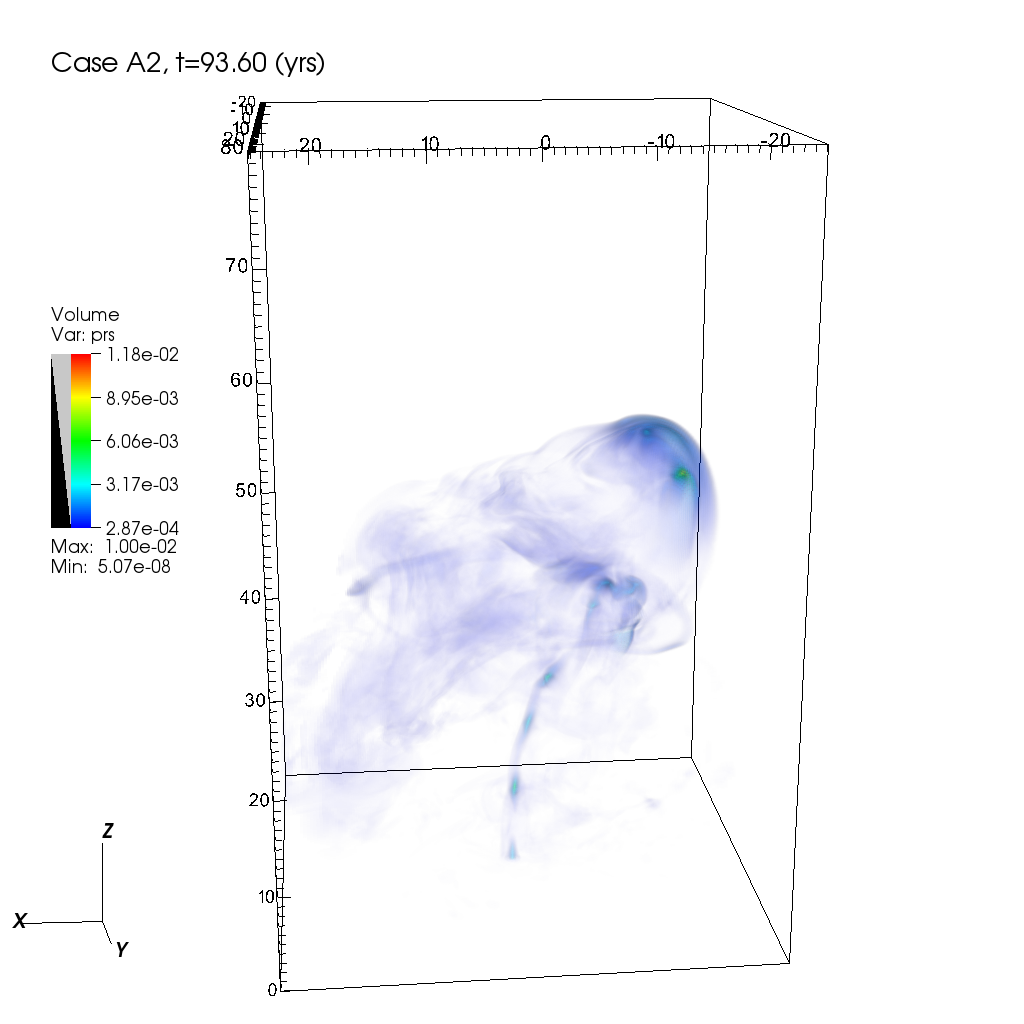}
  \includegraphics*[width=0.32\textwidth]{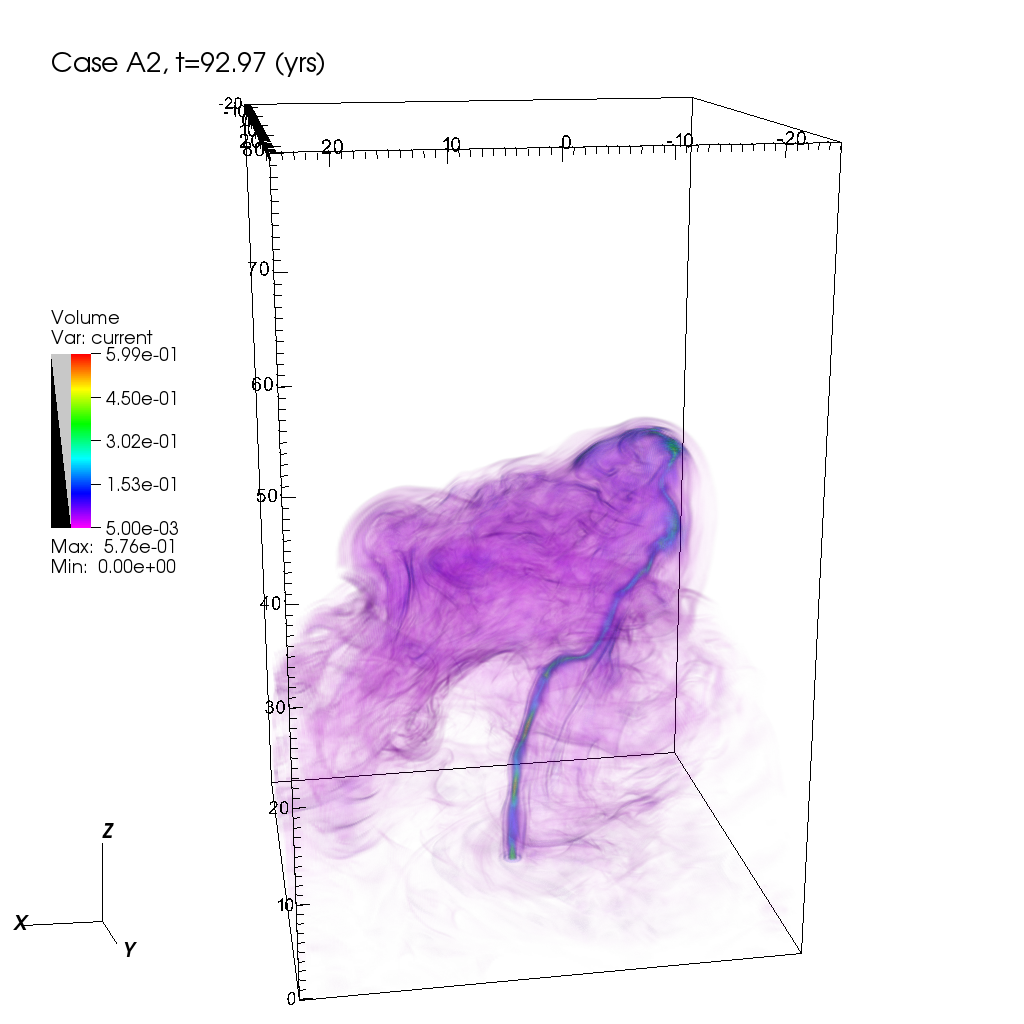}%
  \includegraphics*[width=0.32\textwidth]{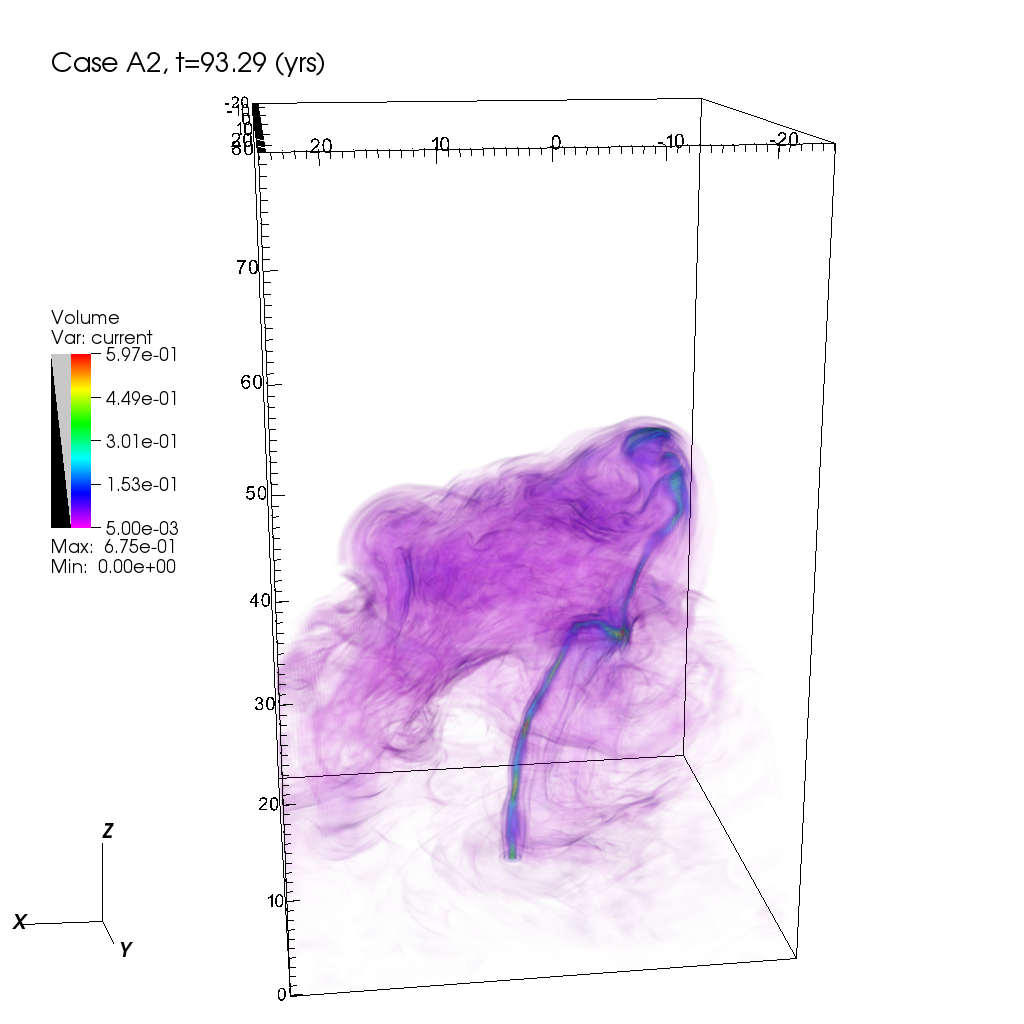}%
  \includegraphics*[width=0.32\textwidth]{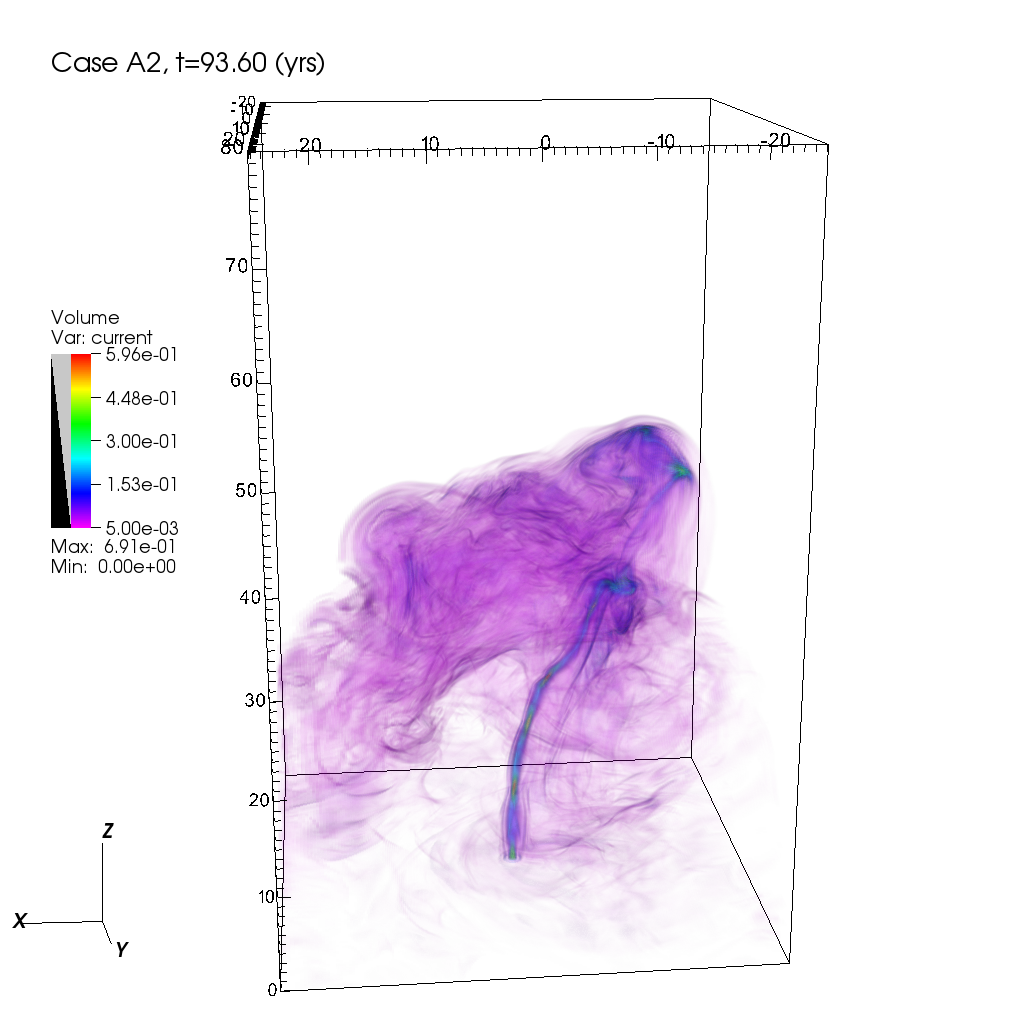}
  \includegraphics*[width=0.32\textwidth]{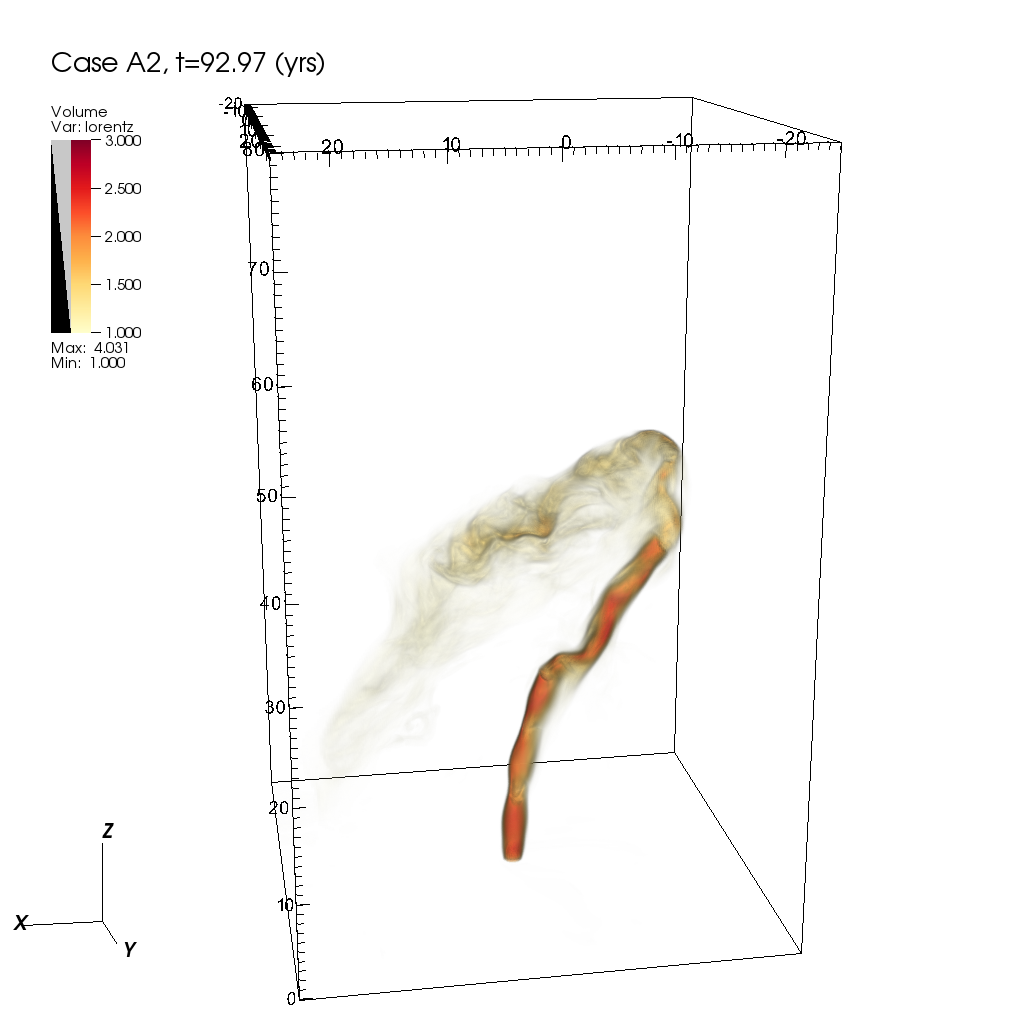}%
  \includegraphics*[width=0.32\textwidth]{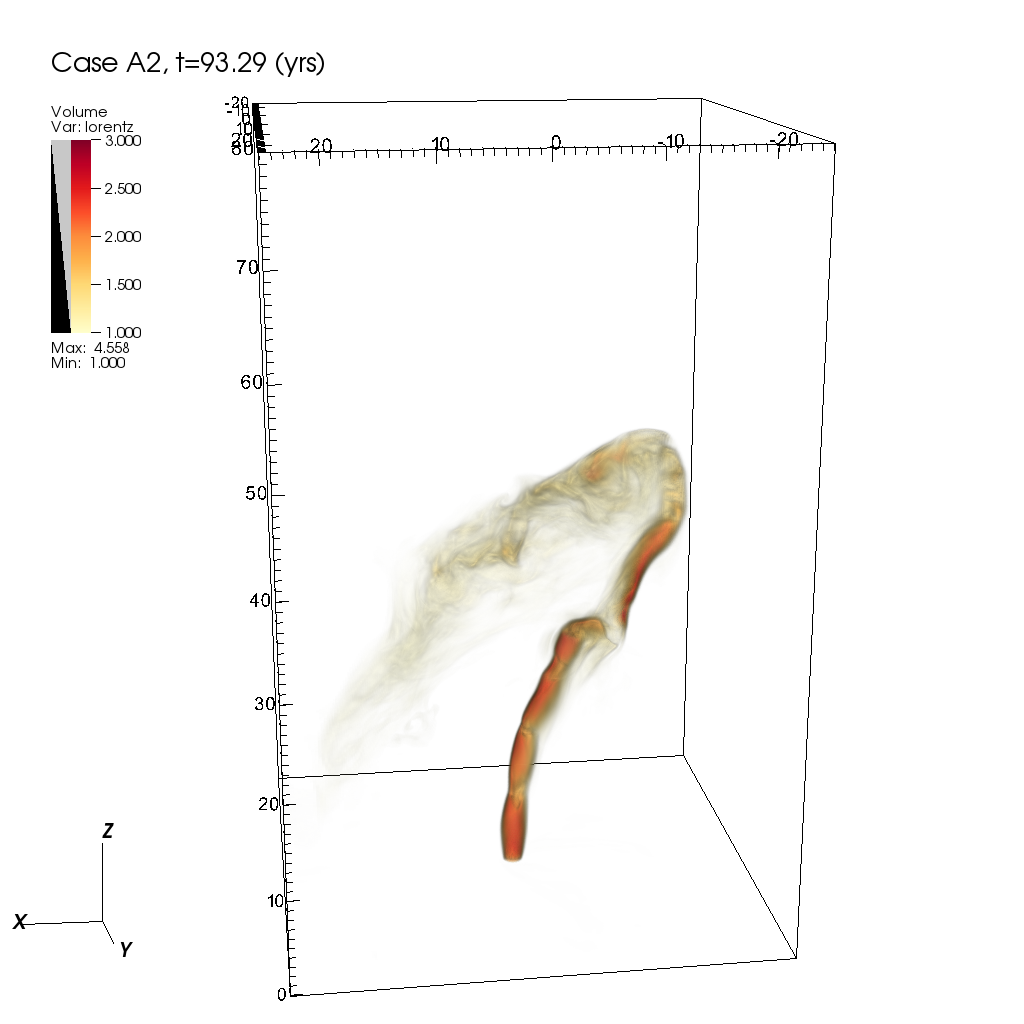}%
  \includegraphics*[width=0.32\textwidth]{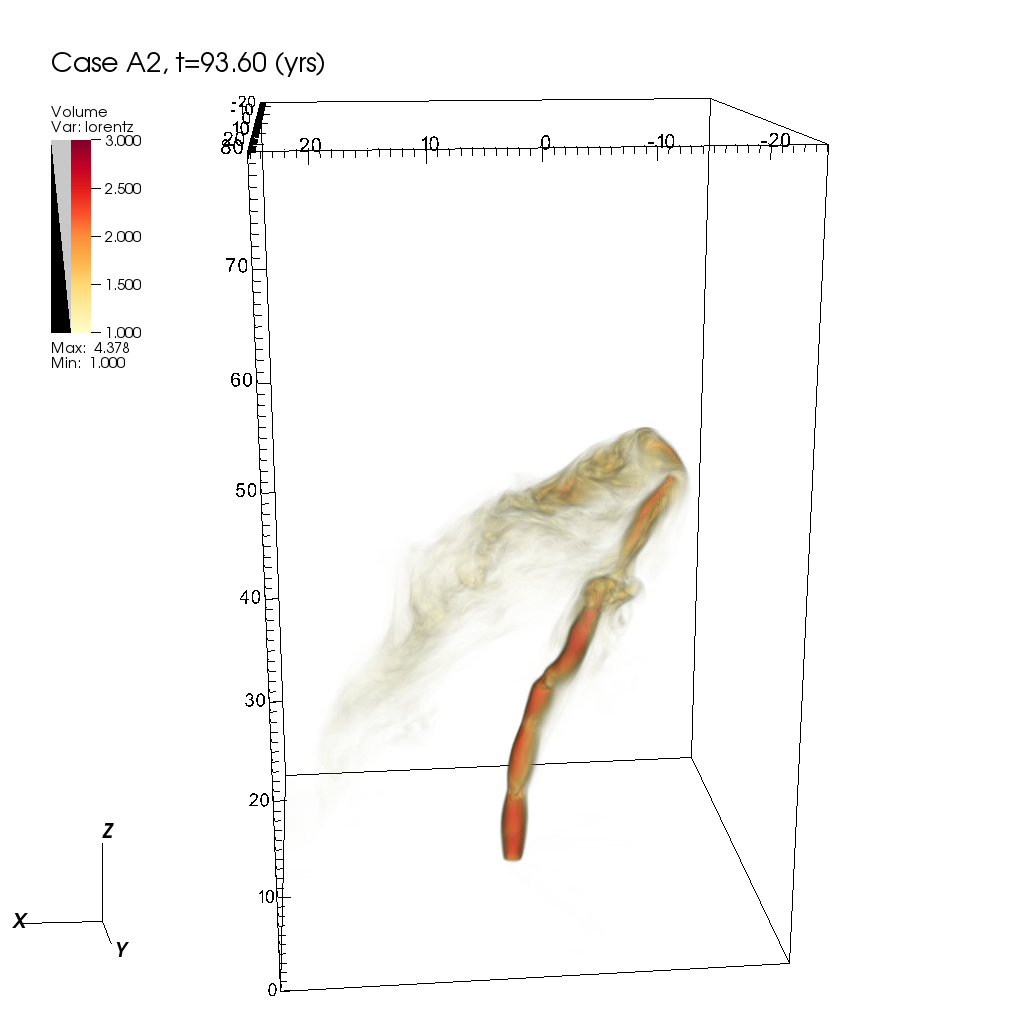}
  \caption{\small Jet fragmentation for the A2 case illustrated through a sequence of close-by frames at $t\approx 93$ (left column), $t\approx 93.3$ (central column) and $t\approx 93.6$ (right column) years.
  From top to bottom the figure shows, respectively, the volume rendering of the $\sigma$ parameter, thermal pressure, current density magnitude $|\vec{J}|=|\nabla\times\vec{B}/(4\pi)|$ and bulk flow Lorentz factor.}
  \label{fig:A2-fragmentation}
\end{figure*}

As the jet advances into the remnant, the propagation is accompanied by the formation of highly intermittent unstable structures during which jet fragmentation is frequently observed.
These events take place on a short time-scale (typically less than a year) and
in correspondence of large kinked deflection where the jet beam temporarily breaks down forming strong intermediate shock waves resembling the main termination shock.
A typical example is illustrated in Fig. \ref{fig:A2-fragmentation} where we show, from left to right, a volume rendering of the $\sigma$ parameter, the current density and the Lorentz factor for the A2 jet during a short temporal evolution.
 
We point out that the features produced in our jet fragmentation (hot-spots) do not have an equivalent observed optical or X-ray signature, although some hints of such structures are present in the X-ray maps of the jet terminal part (see left panel in Fig. \ref{fig:comparison}).
On the other hand, the locations of our jets that correspond to higher values of, e.g, the $\sigma$ parameter, might have a relation with the synchrotron processes responsible for the optical and X-ray emission. 
This issue will be further investigated in forthcoming studies.


\subsubsection{Magnetic Field Topology.}
%

\begin{figure}
 \begin{center}
  \includegraphics[width=0.45\textwidth]{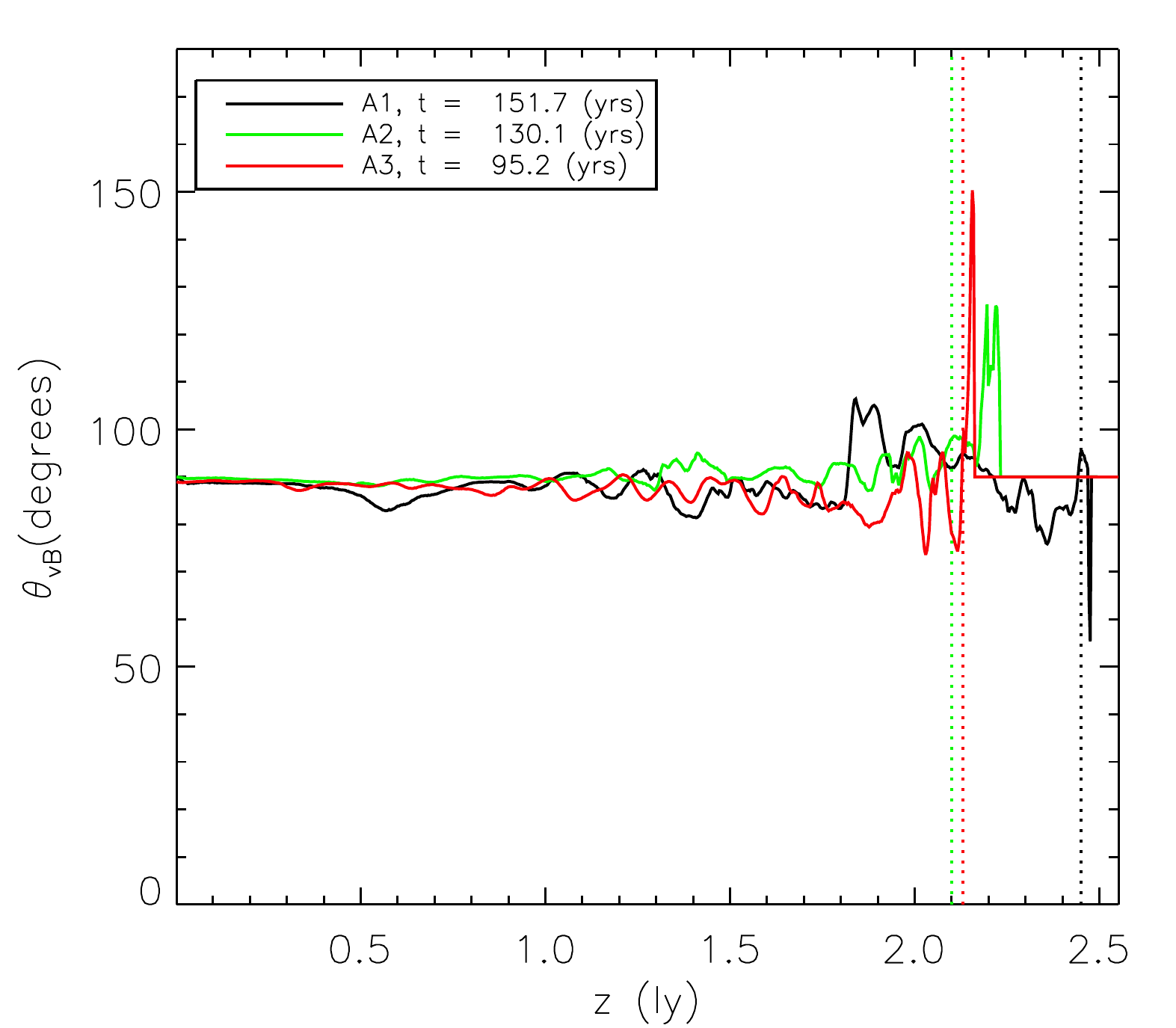}
  \caption{\small Average angle between the magnetic field vector and the mean flow direction as a function of $z$ for cases A1, A2 and A3 (solid lines).
  The dotted lines show the (approximate) position of the terminal reverse shock for the three cases.
  The evolution times corresponds to those indicated in Fig. \ref{fig:final_snapshots}}
  \label{fig:cvb}
 \end{center}
\end{figure}
\begin{figure}
 \begin{center}
  \includegraphics*[width=0.45\textwidth]{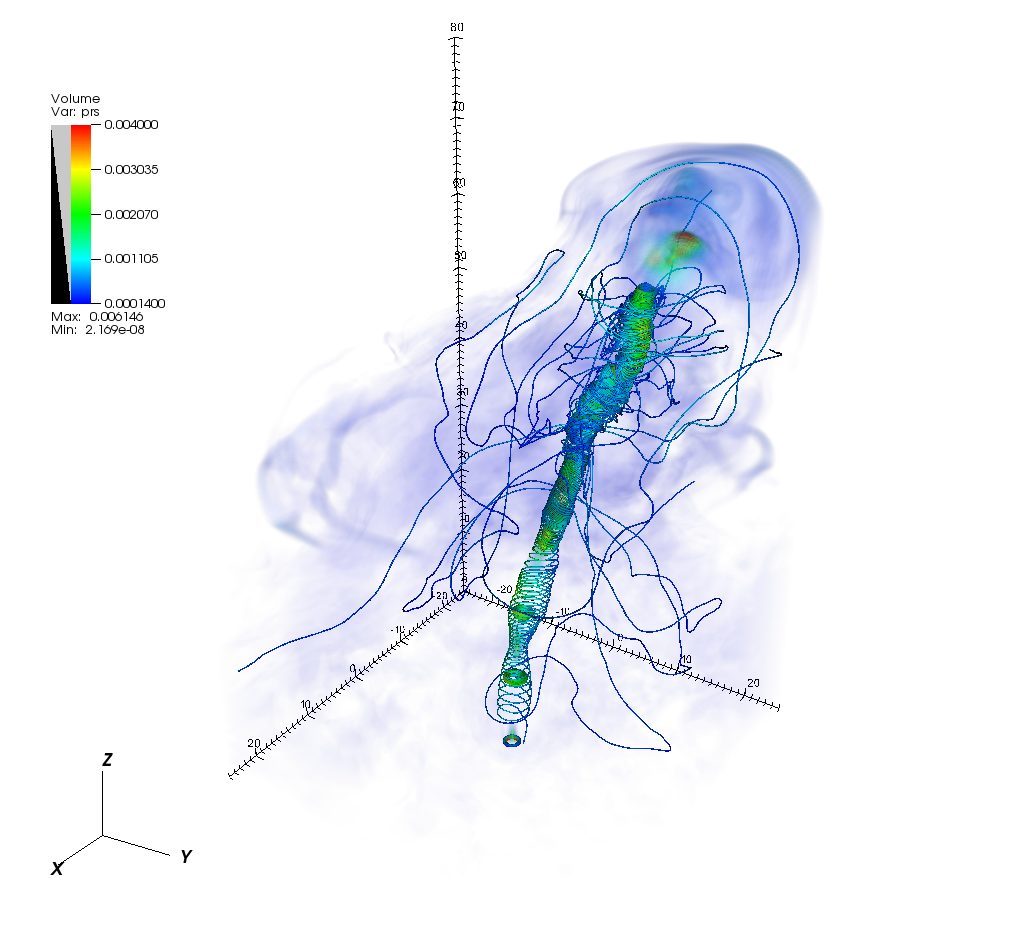}
  \caption{\small  Magnetic field lines and pressure volume map for the A2 jet at $t\sim 130$ years.}
  \label{fig:A2.field_lines}
 \end{center}
\end{figure}

In order to gain some insight on the topology of magnetic field, we first compute the average flow direction by integrating, for each $z$, the velocity vector on horizontal planes:
\begin{equation}\label{eq:average_velocity}
  \hvec{n}(z) = \frac{\left<\vec{v},\chi_j\right>}
                     {|\left<\vec{v},\chi_j\right>|}\,,
\end{equation}
with $\chi_j$ defined by Eq. (\ref{eq:jet_chi}).
We then decompose the magnetic field into components that are parallel and perpendicular to the direction given by $\hvec{n}$:
\begin{equation}\label{eq:Bsplit}
  \vec{B} = B_\parallel\hvec{n}(z) + \vec{B}_\perp \,,
\end{equation}
where $B_\parallel = \vec{B}\cdot\hvec{n}(z)$ is the magnetic field component parallel to the horizontally-averaged velocity vector and $\vec{B}_\perp$ is the component of the field perpendicular to it.
The average cosine between magnetic field and mean flow direction is then simply obtained from:
\begin{equation}
  \av{\theta}_{vB} = \mathrm{acos}\left< \frac{B_\parallel}{|\vec{B}|},\chi_j\right> \,.
\end{equation}

Fig. \ref{fig:cvb} shows that the magnetic field in the jet remains essentially  perpendicular to the (average) flow trajectory for most of the jet length while it acquires a poloidal component immediately after the terminal reverse shock.
This is confirmed by the direct three-dimensional visualization of the magnetic field lines in proximity of the beam (shown in Fig. \ref{fig:A2.field_lines} for the A2 jet) which reveals that the field retains the initial toroidal shape that  progressively evolves into a bent helical structure with a small pitch.
Indeed, as mentioned in section \ref{sec:internal_struct}, the magnetic field acts as an effective screening sheath that hampers the development of small scale perturbations at the interface between the jet and the surrounding.

\subsection{Jet Deflections.}
\label{sec:deflection}
%

\begin{figure}
 \begin{center}
  \includegraphics[width=0.5\textwidth]{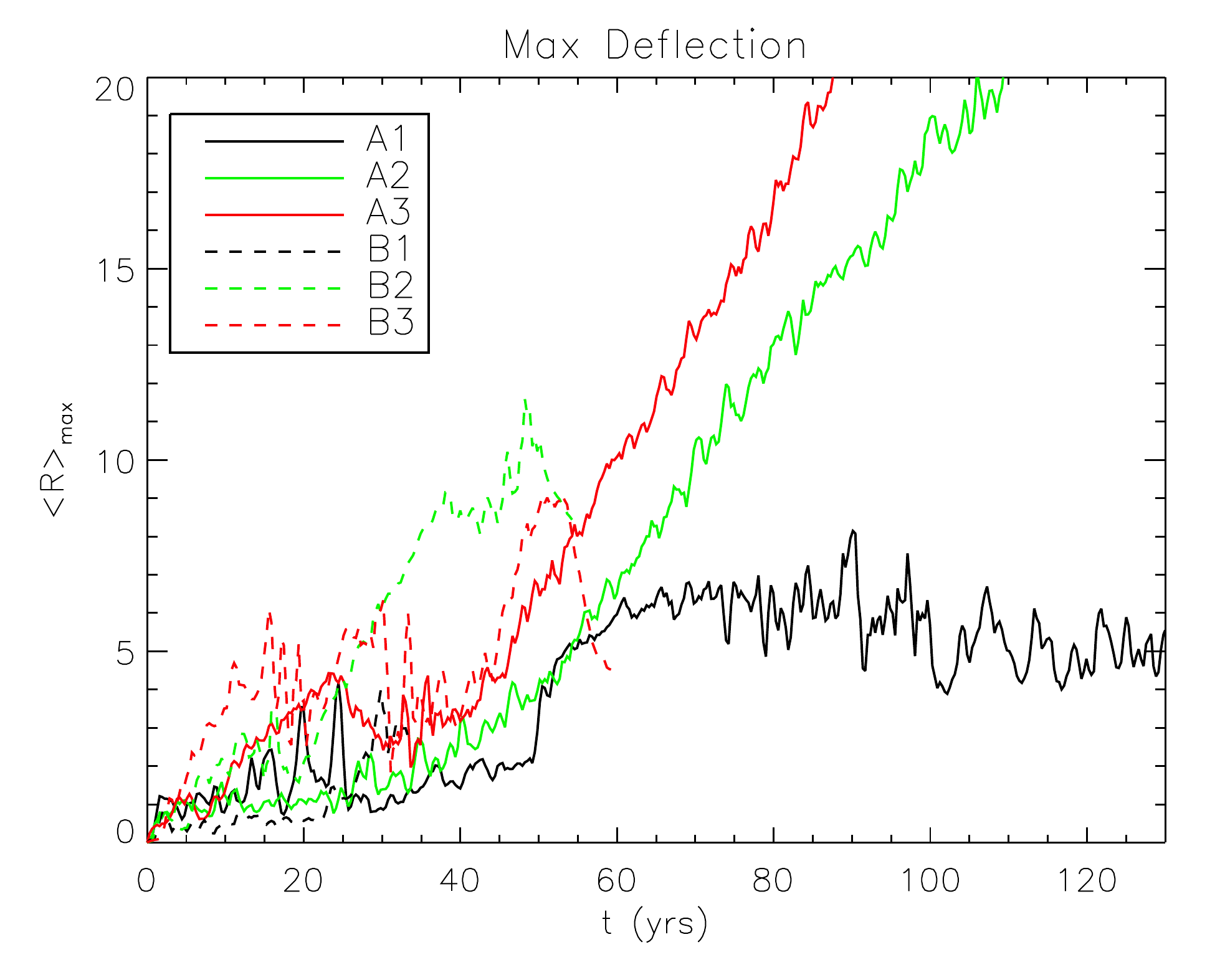}
  \caption{\small Maximum radial deflection as a function of time for the different cases discusses in the text.
  Solid lines refer to case A1 (black), case A2 (green) and A3 (red) while dashed lines refer to case B1, B2 and B3, respectively.}
  \label{fig:cm_max}
 \end{center}
\end{figure}

A remarkable feature observed in several runs is a curved trajectory featuring large time-dependent deflection of the jet beam away from the main longitudinal $z$ axis (see Fig. \ref{fig:final_snapshots}).
As discussed by \cite{MRBFM.2010}, this behavior may be ascribed to the onset of CD instabilities triggered by the presence of a toroidal magnetic field component
\citep[see also][]{MSO.2008, Porth.2013}.
This result is confirmed by numerical investigations of infinitely-long periodic jets adopting the same initial structure (Mignone et al., in preparation) revealing  the presence of CD instabilities with a rapid growth of the $m=1$ (or kink) mode.

We introduce a measure of the deflection radius $\av{R}(t,z) = \sqrt{\av{x}^2(t,z) + \av{y}^2(t,z)}$ and the deflection angle $\av{\phi}(t,z) = \tan^{-1}(\av{y}(t,z)/\av{x}(t,z))$ by computing at each time and vertical height the centroids $\av{x}(t,z)$ and $\av{y}(t,z)$:
\begin{equation}\label{eq:centroids}
  \av{x}(t,z) = \left<x,\chi_j\right>
   \,,\qquad
  \av{y}(t,z) = \left<y,\chi_j\right> \,,
\end{equation}
and choosing the weight function $\chi_j$ as in Eq. (\ref{eq:jet_chi}).

Fig. \ref{fig:cm_max} plots the maximum of $\av{R}(t,z)$ taken over the spatial coordinate $z$ as a function of time. 
Case A2 and A3 show the largest bendings reaching values in excess of $\approx 20$ jet radii.
Although to a less degree, cases B2 and B3 are also prone to appreciable wiggling ($\av{R}_{\max} \approx 10$) whereas case B1 propagates almost parallel to the main longitudinal axis with very weak bending of the beam.

\begin{figure}
 \begin{center}
  \includegraphics[width=0.5\textwidth]{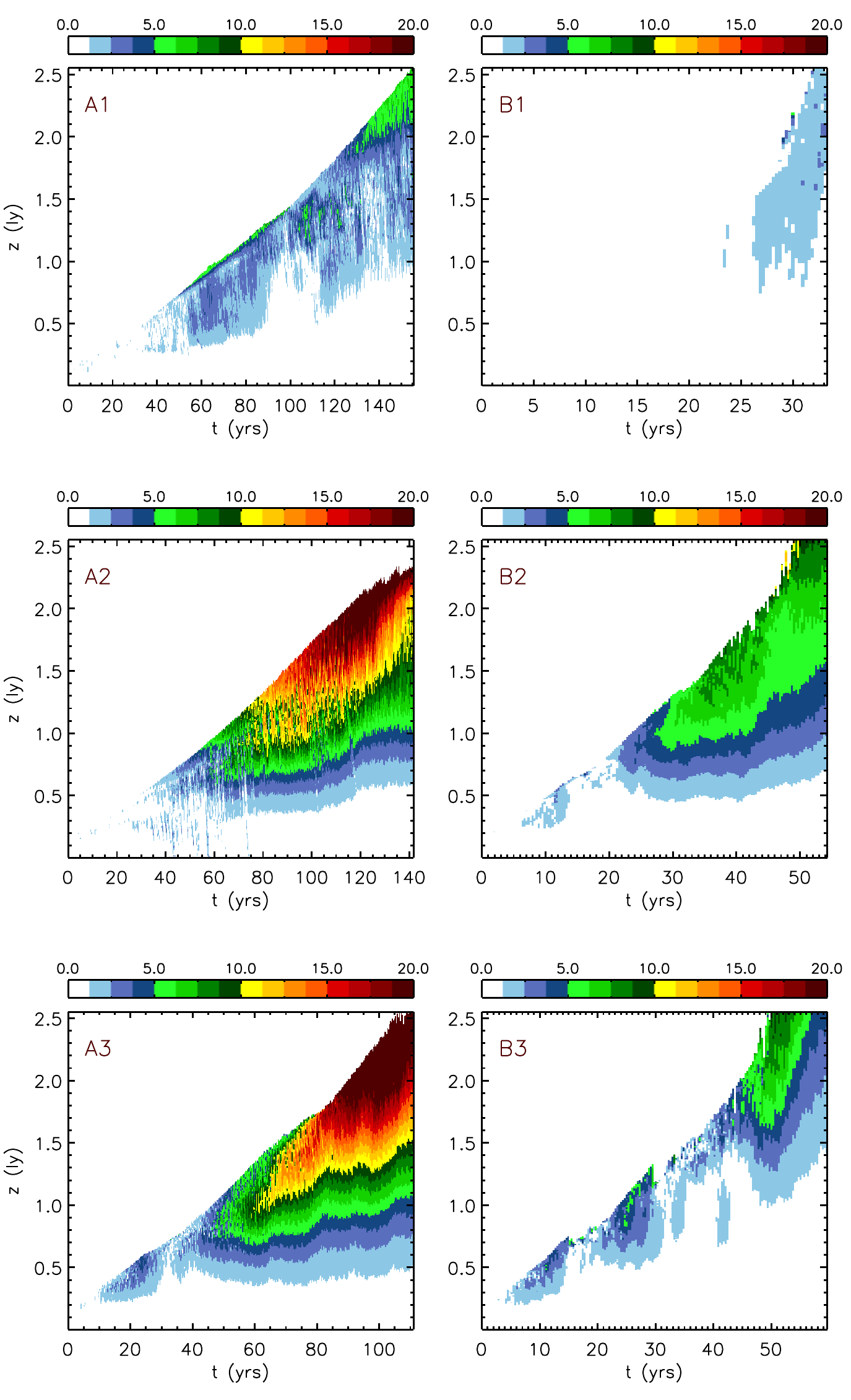}
  \caption{\small Two-dimensional colored distribution maps of the average deflection radius $\av{R}$ (in units of the jet radius) as function of time (abscissa, in years) and vertical height (ordinate, in light-years) for the six simulated cases.
  The amount of deflection at a given time $t$ and height $z$ is quantified by a different color given by the varied shades of blue ($\av{R}\in[0,5]$), green ($\av{R}\in[5,10]$), orange ($\av{R}\in[10,15]$) and red ($\av{R}\in[15,20]$).}
  \label{fig:barycenter-rho-rad}
 \end{center}
\end{figure}

\begin{figure}
 \begin{center}
  \includegraphics[width=0.5\textwidth]{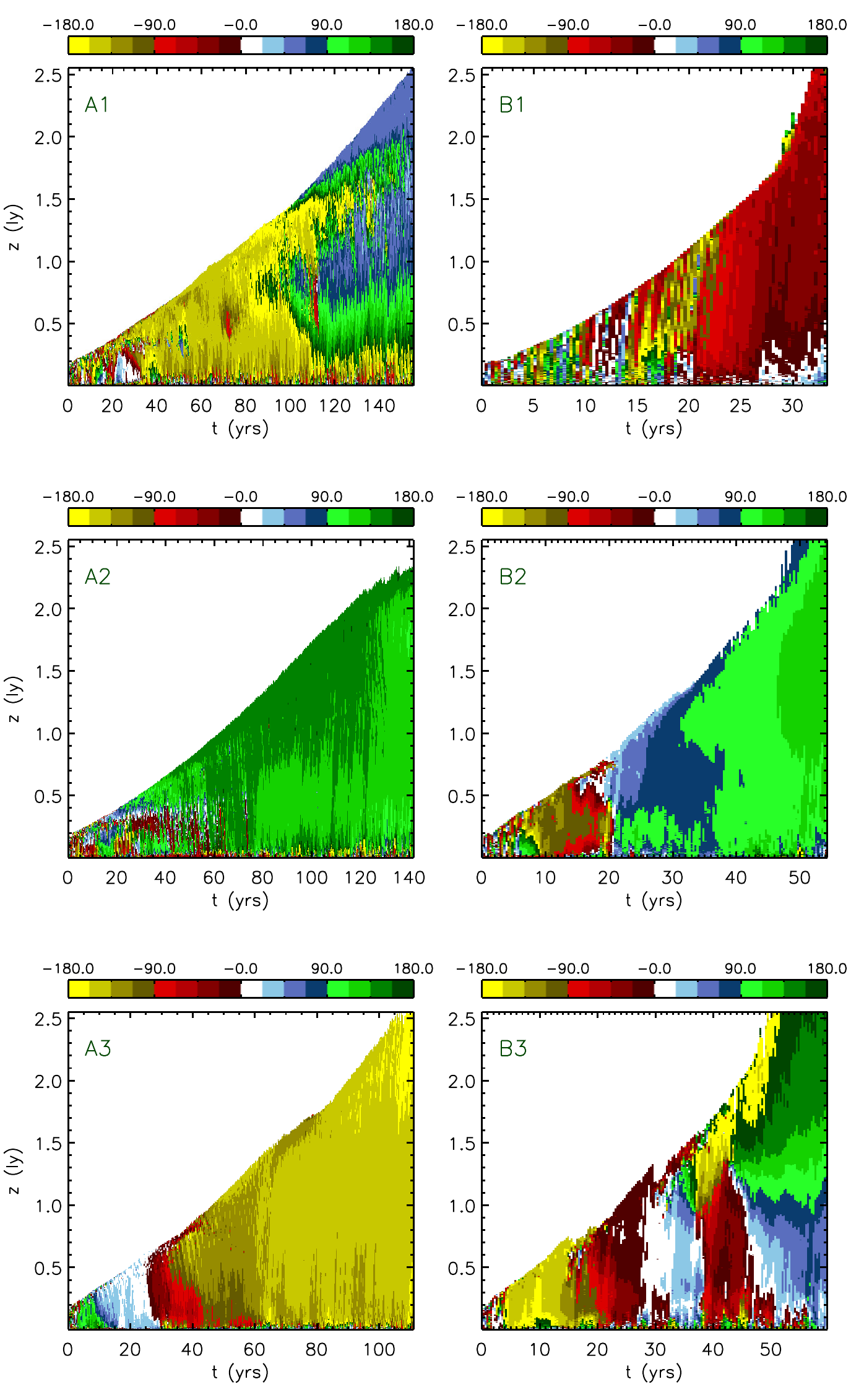}
  \caption{\small Two-dimensional colored distribution maps (similar to Fig. \ref{fig:barycenter-rho-rad}) of the average deflection angle $\av{\phi}$ as function of time (abscissa, in years) and vertical height (ordinate, in light-years) for the six simulated cases.}
  \label{fig:barycenter-rho-phi}
 \end{center}
\end{figure}

\begin{figure}
 \begin{center}
  \includegraphics[width=0.45\textwidth]{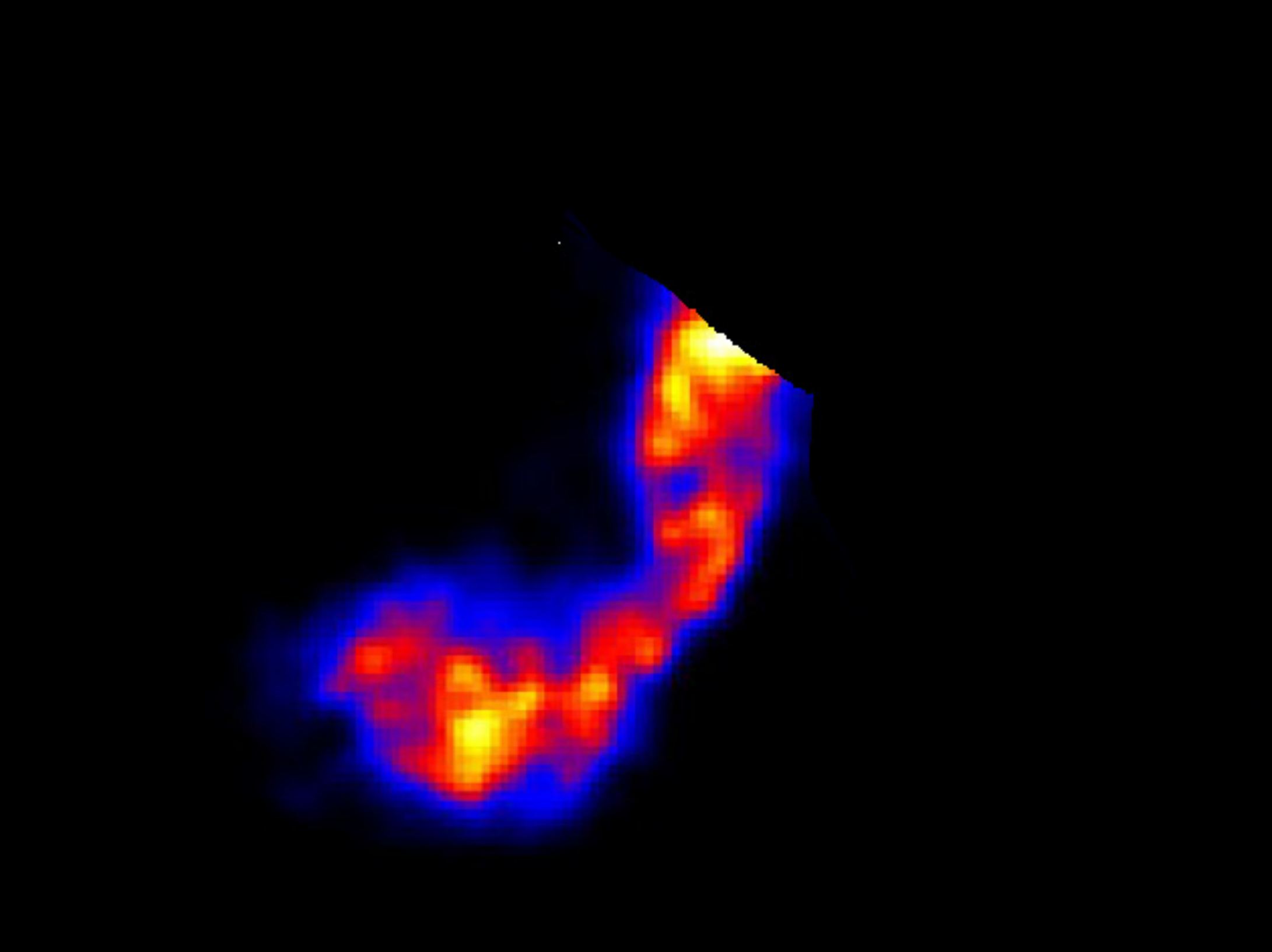}
  \includegraphics[width=0.45\textwidth]{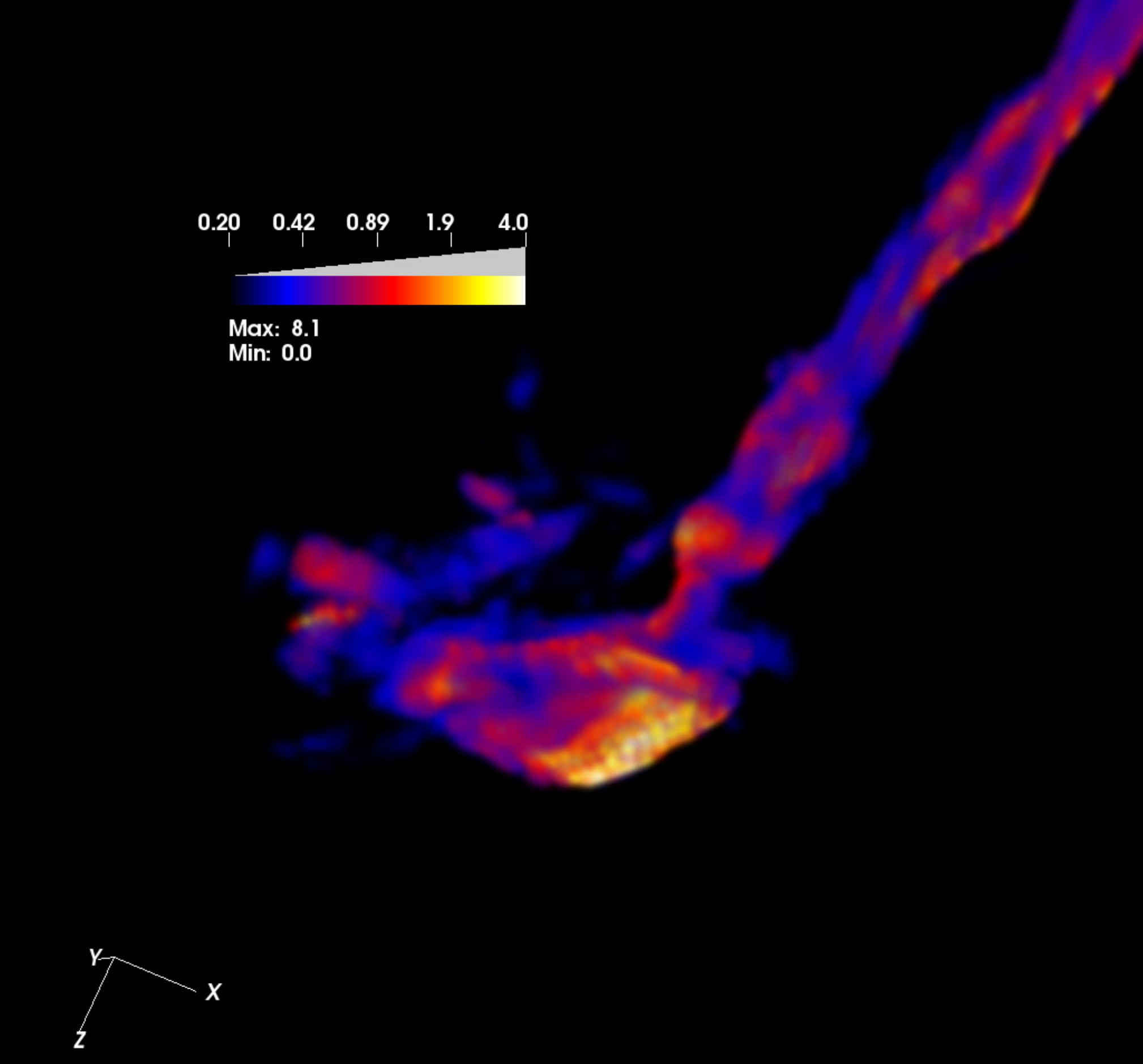}
  \caption{\small An example of qualitative agreement between data (top showing Chandra X-ray map of the terminal jet part) and simulations (bottom, showing a volume rendering of $\log\sigma$ for the A2 jet).}
  \label{fig:comparison} 
 \end{center}
\end{figure}

A more detailed investigation is given in Fig. \ref{fig:barycenter-rho-rad} and Fig. \ref{fig:barycenter-rho-phi} showing 2D colored distribution maps of the deflection radius $\av{R}$ and angle $\av{\phi}$ as functions of time and vertical coordinate.
Jets become increasingly more flexed as they propagate from the injection region up to the head where they attain the largest deformations \citep{Porth.2013}.
The shape of these deformed structures roughly resembles the morphological features that can be inferred from observational data. 
This is shown in Fig. \ref{fig:comparison} where we present a qualitative comparison between the X-ray Chandra observation of the terminal part of the jet in 2010 (top panel) and our simulated jet (case A2, bottom panel).


For moderate and strong magnetizations in low-velocity jets (A2 and A3), the deflection keeps growing indefinitely during the course of the simulation.
In these cases we observe, for $t \lesssim 50\,{\rm yrs}$, a change in the propagation direction on a time scale of $\sim 10$ years while, afterwards, jet material keeps flowing along a particular curved direction established by the large-scale backflow circulation pattern and changes direction very slightly.
Here the inertia of the flexed jet becomes so effectively large that any restoring mechanism is unable to push the beam back along the main longitudinal axis.
In the corresponding higher velocity cases (B2 and B3) the amount of bending is reduced owing to the increased inertia which acts as a stabilizing factor.
In these cases the jet changes its propagation angle by a large amount ($\sim 180^\circ$ or more) around the main longitudinal axis over a time period $\lesssim 40$ years (see the right panels in Fig. \ref{fig:barycenter-rho-phi}).

We point out, however, that previous simulations (not shown here) have demonstrated that the evolution of the trajectory and the corresponding amount of deflection may be noticeably affected by the shape of the initial perturbation (given by Eq. \ref{eq:jet_perturbation}).


We thus conclude that the magnetization parameter $\sigma$ plays a crucial role in destabilizing the jet: weakly magnetized configurations (A1 and B1) are less affected by the growth of instability than the corresponding moderately (A2 and B2) and highly (A3 and B3) magnetized models which show comparable growth rates.
The effect of the Lorentz factor, on the other hand, is that of reducing the growth of instability, in accordance with the results obtained from the linear stability analysis of \cite{Bodo_etal.2013}.

\subsubsection{Flow Direction.}
%

\begin{figure}
 \begin{center}
  \includegraphics[width=0.45\textwidth]{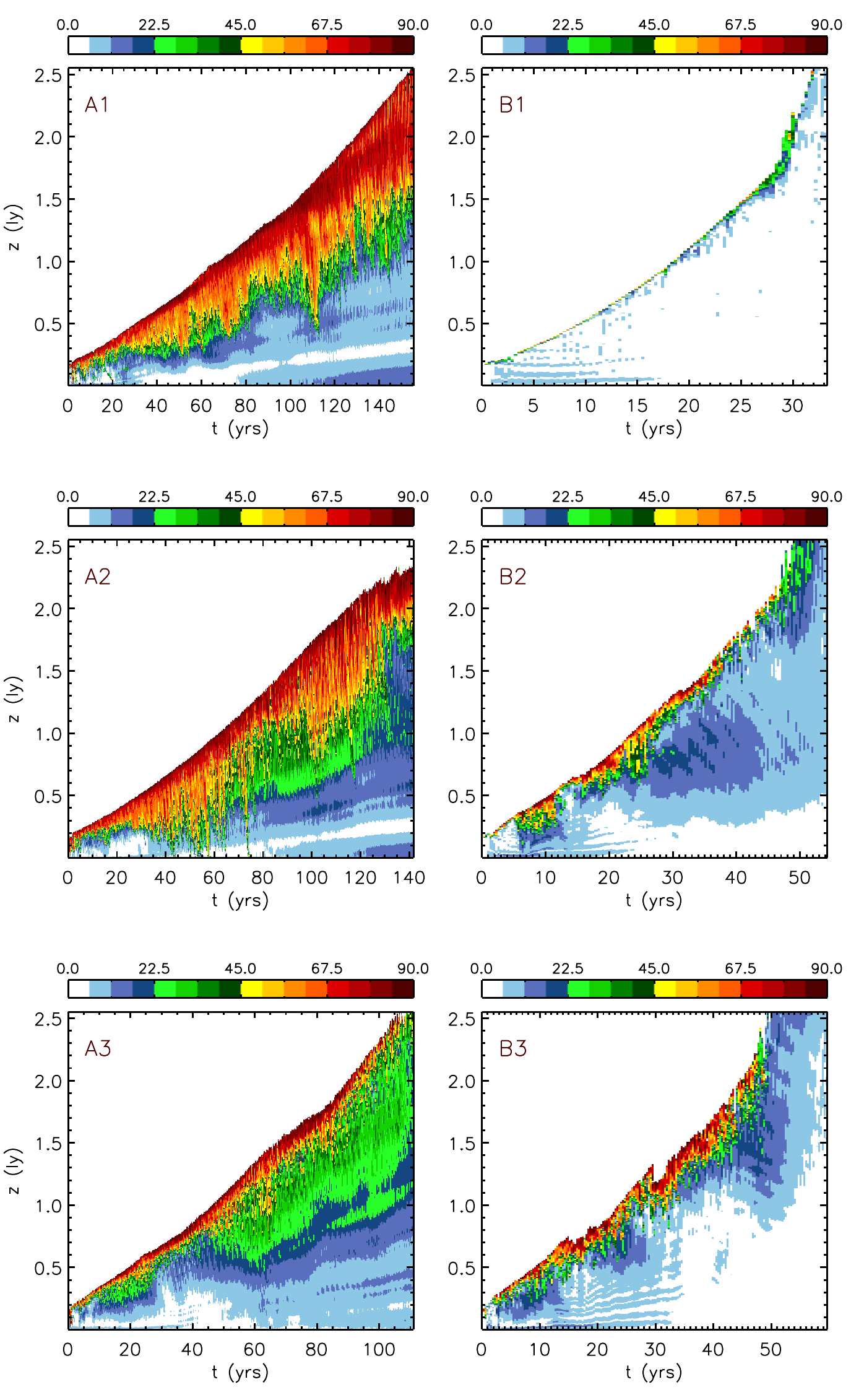}
  \caption{\small Same as Fig. \ref{fig:barycenter-rho-rad} but for the average propagation angle $\av{\theta}_+$ (in degrees) in the positive $z$ direction.
  The color map spans from $0^\circ$ (white) to $90^\circ$ (dark red).}
  \label{fig:acosp} 
 \end{center}
\end{figure}
\begin{figure}
 \begin{center}
  \includegraphics[width=0.5\textwidth]{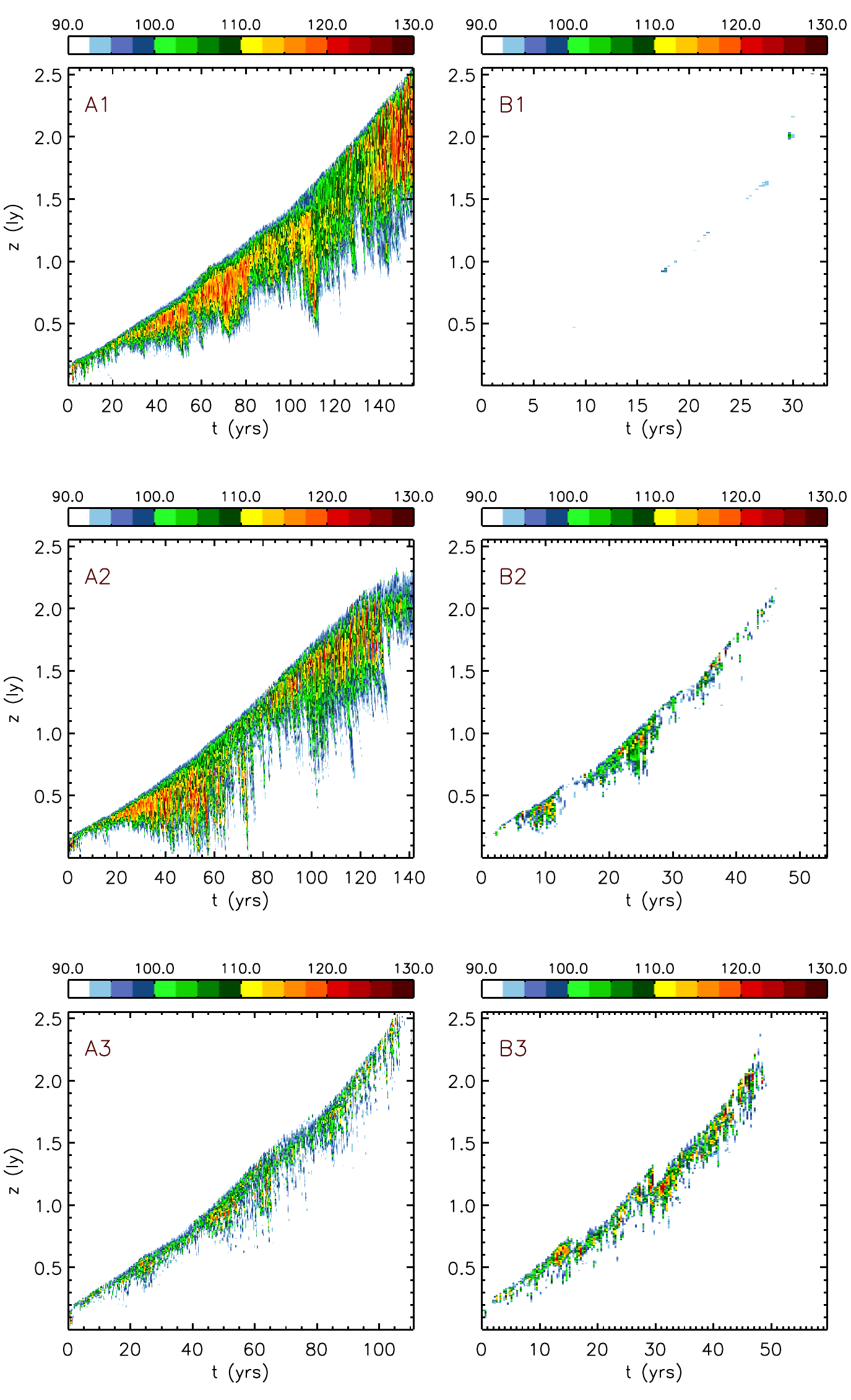}
  \caption{\small Same as Fig. \ref{fig:acosp} but for the negative $z$ direction.
  In this case, $\av{\theta}_-$ ranges from $90^\circ$ (white) to $180^\circ$ (dark red).}
  \label{fig:acosm}
 \end{center}
\end{figure}

The change in the jet trajectory is associated with a corresponding variation of the average propagation velocity.
We compute the average angle $\av{\theta}$ between the velocity vector and the axial direction by integrating, on constant $z$-planes, the projected velocity:
\begin{equation}\label{eq:theta_pm}
  \av{\theta}_{\pm} = \mathrm{acos}\left<\frac{v_{z,\pm}}{|\vec{v}|},\chi_j\right> \,,
\end{equation}
where $v_{z,+} = \max(\vec{v}\cdot\hat{\vec{e}}_z,0)$ and $v_{z,-} = \min(\vec{v}\cdot\hat{\vec{e}}_z,0)$ are used to discriminate between jet material moving in the positive or negative vertical direction, respectively.
In other words, $\av{\theta}_+\in[0,90^\circ]$ gives a measure of the forward flow while $\av{\theta}_-\in[90,180^\circ]$ is associated with the backflow motion.
The filter function is defined as usual (Eq. \ref{eq:jet_chi}).

Fig. \ref{fig:acosp} shows a colored distribution map of $\av{\theta}_+$ as a function of time and vertical distance.
In all cases, sudden changes in the trajectory occur at the jet head where magnetic field are amplified and the flow is abruptly decelerated through the termination shock.
However, low-speed jets tend to assume a large-scale curved structured since the average propagation angle gradually changes from $\approx 0^\circ$ (straight propagation) close to the launching region up to $\approx 90^\circ$ at the jet head.
Conversely, high speed jets are stabilized by the larger Lorentz factor and propagate more parallel to the longitudinal axis building large kicks mainly in proximity of the jet head.
Here the velocity is drastically reduced and the magnetic field is increased thus de-stabilizing the motion.

\subsubsection{Backflow Motion.}
%

\begin{figure*}
 \begin{center}
  \includegraphics*[width=0.32\textwidth]{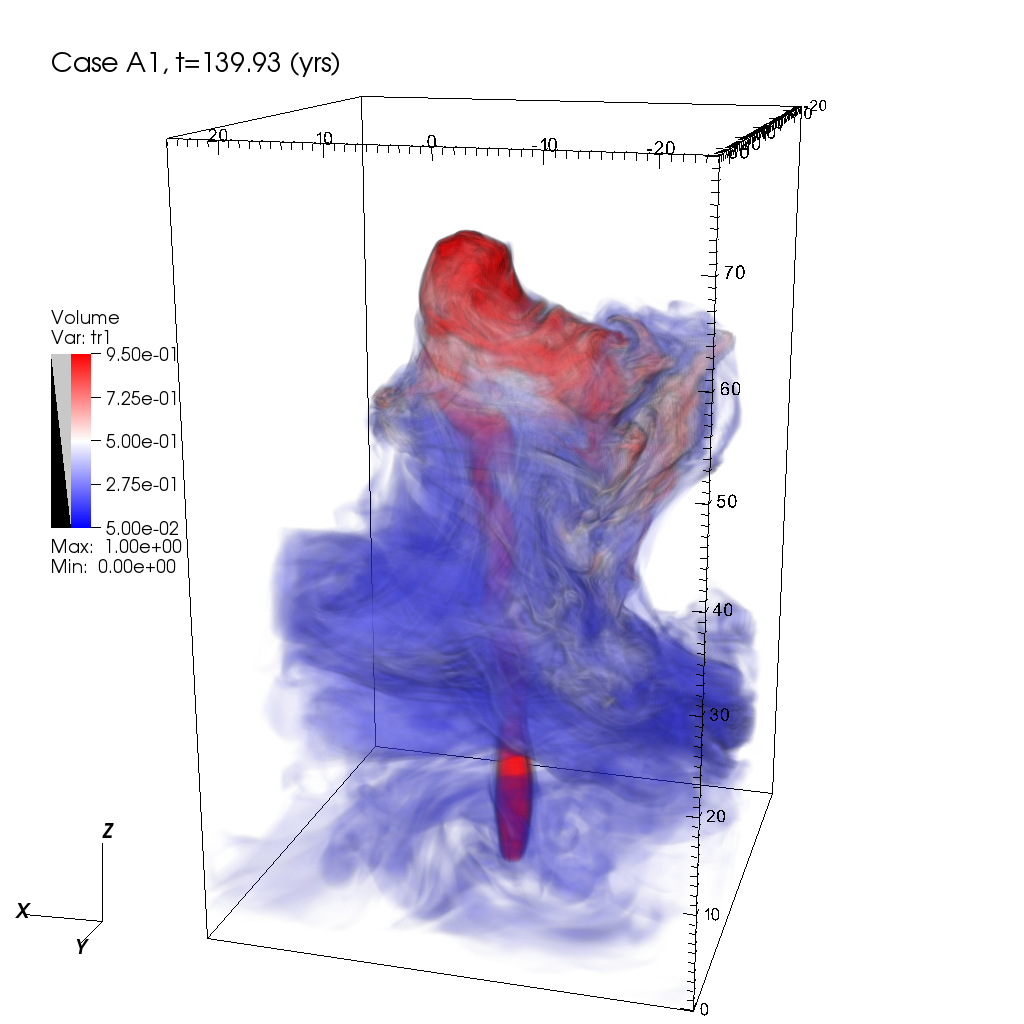}
  \includegraphics*[width=0.32\textwidth]{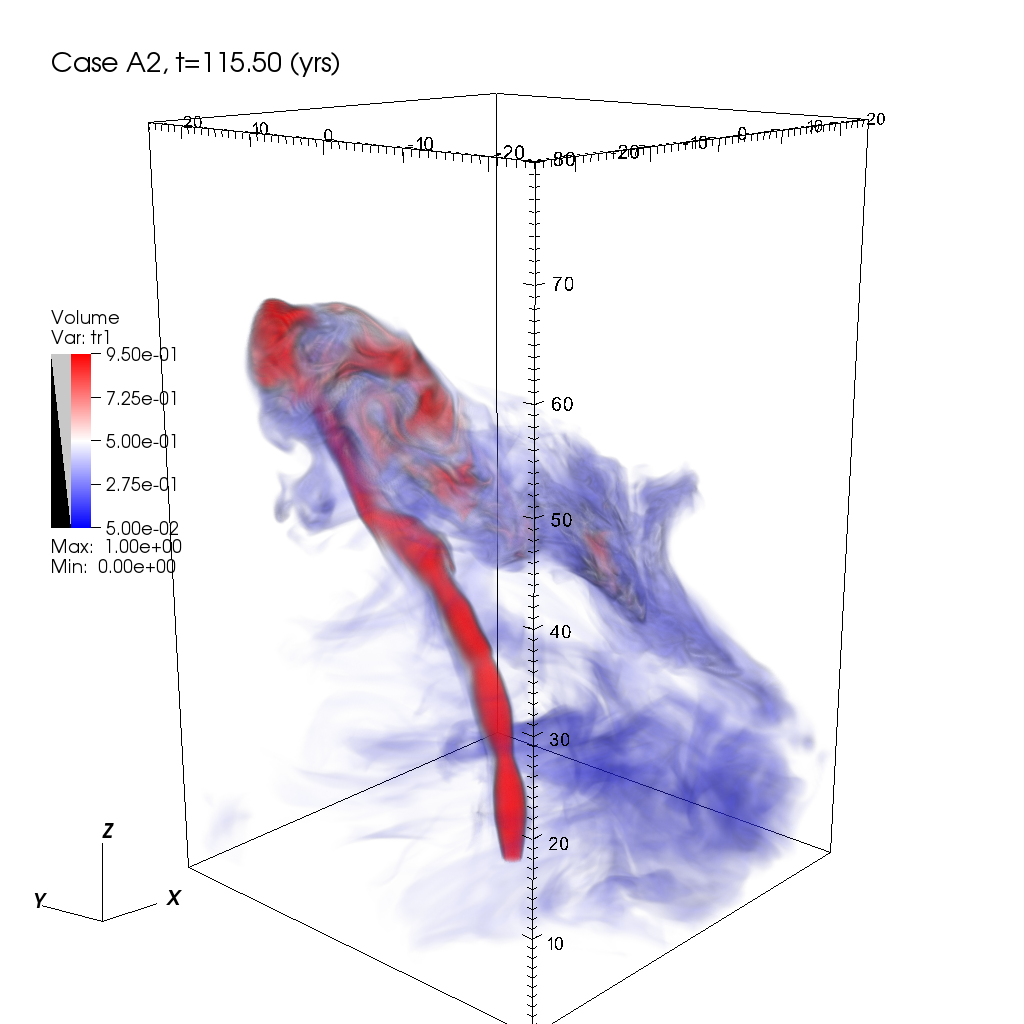}
  \includegraphics*[width=0.32\textwidth]{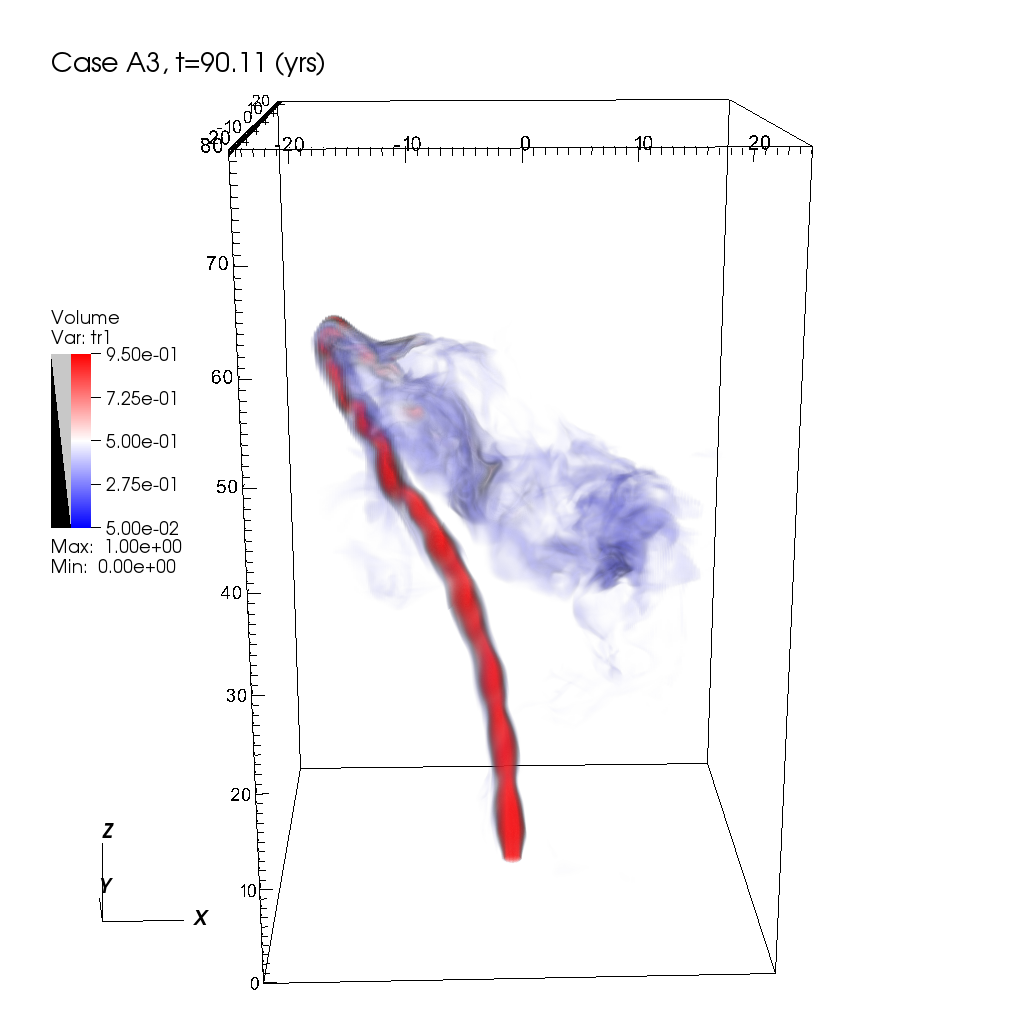}
  \caption{\small Three-dimensional rendering of the passive scalar distribution for the A1, A2 and A3 jets at the time reported in the legend.
  The size of the backflow region reduces with increasing magnetization.
  Regions in red mark fluid elements composed by at least $50\%$ of the jet material (${\cal T} > 0.5$) whereas blue regions mark fluids elements that have mixed with remnant and are composed by less than $50\%$ of the jet material (${\cal T} < 0.5$).}
  \label{fig:A.tracer}
 \end{center}
\end{figure*}

The departure from axial symmetry produced by kinked deformations has the side effect of promoting a predominantly one-sided backflow motions along the negative $z$ direction.
This is shown in Fig. \ref{fig:acosm} where we plot a 2D color map of $\av{\theta}_-$ (computed from Eq. \ref{eq:theta_pm}) as function of time and vertical height.
Backflows are strongest in case A1 and progressively lessen at larger magnetizations or for larger Lorentz factor cases (e.g., they are almost absent in the B1 jet).
A three-dimensional view is given in Fig. \ref{fig:A.tracer} for the small Lorentz factor jets.
In the most prominent cases, the backflow is able to survive over a distance of $\approx 1 \,{\rm ly}$ and for several years before its kinetic energy is dissipated in the form of turbulent motion during the interaction with the ambient medium.

\subsection{Dissipation.}
%

A crucial aspect in the modeling of magnetically driven outflows is the dissipation of magnetic fields at both large and small scales.
This is discussed in the following sections.

\subsubsection{Large-scale Dissipation}
%

\begin{figure}
  \includegraphics*[width=0.45\textwidth]{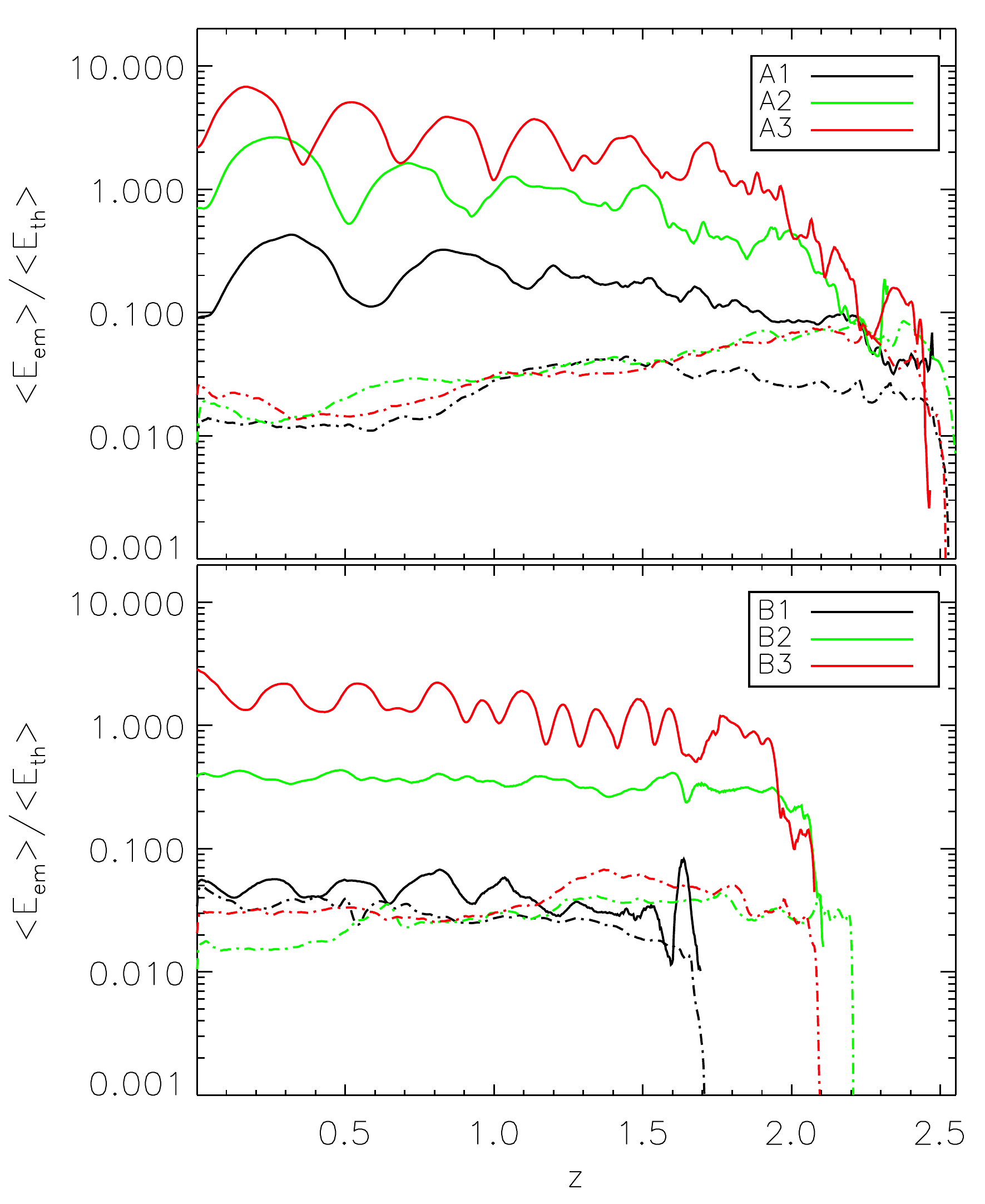}
  \caption{\small Ratio between average electromagnetic and thermal energies inside the jet (solid lines) and outside in the cocoon (dot-dash).
  Plots in the top and bottom panels refers to the slower and faster jet cases, respectively.}
  \label{fig:em_et}
\end{figure}

Energy dissipation at large scale may be quantified by comparing the electromagnetic to thermal energy ratios inside the jet with that of the turbulent cocoon.
We distinguish the two regions by taking advantage of the passive scalar ${\cal T}$ and choosing different weight functions.
Inside the jet we compute
\begin{equation}
  \av{E}_{\rm em,j} = \left<\frac{\vec{B}^2 + \vec{E}^2}{8\pi},\chi_j \right>
 \,,\quad
  \av{E}_{\rm th,j} = \left<\frac{p}{\Gamma-1},\chi_j \right>\,,
\end{equation}
with $\chi_j$ defined by Eq. (\ref{eq:jet_chi}).
This choice ensures that only high-velocity regions containing more than $50\%$ of the jet material are included.
On the other hand, in the environment we define
\begin{equation}
  \av{E}_{\rm em,e} = \left<\frac{\vec{B}^2 + \vec{E}^2}{8\pi},\chi_e \right>
 \,,\quad
  \av{E}_{\rm th,e} = \left<\frac{p}{\Gamma-1},\chi_e \right>\,,
\end{equation}
using $\chi_e = 1$ when ${\cal T} < 1/2$, $|\vec{v}| > 0.01$ which includes moving material that has already mixed.
The ratio $\av{E}_{\rm em}/\av{E}_{\rm th}$ is plotted for the jet (solid lines) and the environment (dash-dot lines) in Fig. \ref{fig:em_et} at the end of each simulation case, just before reaching the end of the computational domain.
Inside the jet, the ratio between magnetic and thermal energies present gradually decreasing quasi-periodic oscillations (see \S \ref{sec:internal_struct}) that drop sharply at the jet termination shock.
Conversely, the energy distribution proportions inside the cocoon are more homogeneous and settle down to approximately the same values ($\lesssim 10\%$) independently of the value of jet $\sigma$.
This demonstrates that both Poynting and kinetic energy fluxes are efficiently diverted at the termination shock and thereafter scattered and dispersed via the backflow to feed the cocoon.
During this process the field becomes significantly randomized and the energy distributions reach a state far from equipartition, independently of the initial magnetization.
From this perspective, jets appear to be a very efficient way to dissipate magnetic energy through the interaction of the head of the jet with the surrounding remnant.

\subsubsection{Small-scale Dissipation}
%

\begin{figure}
  \includegraphics*[width=0.45\textwidth]{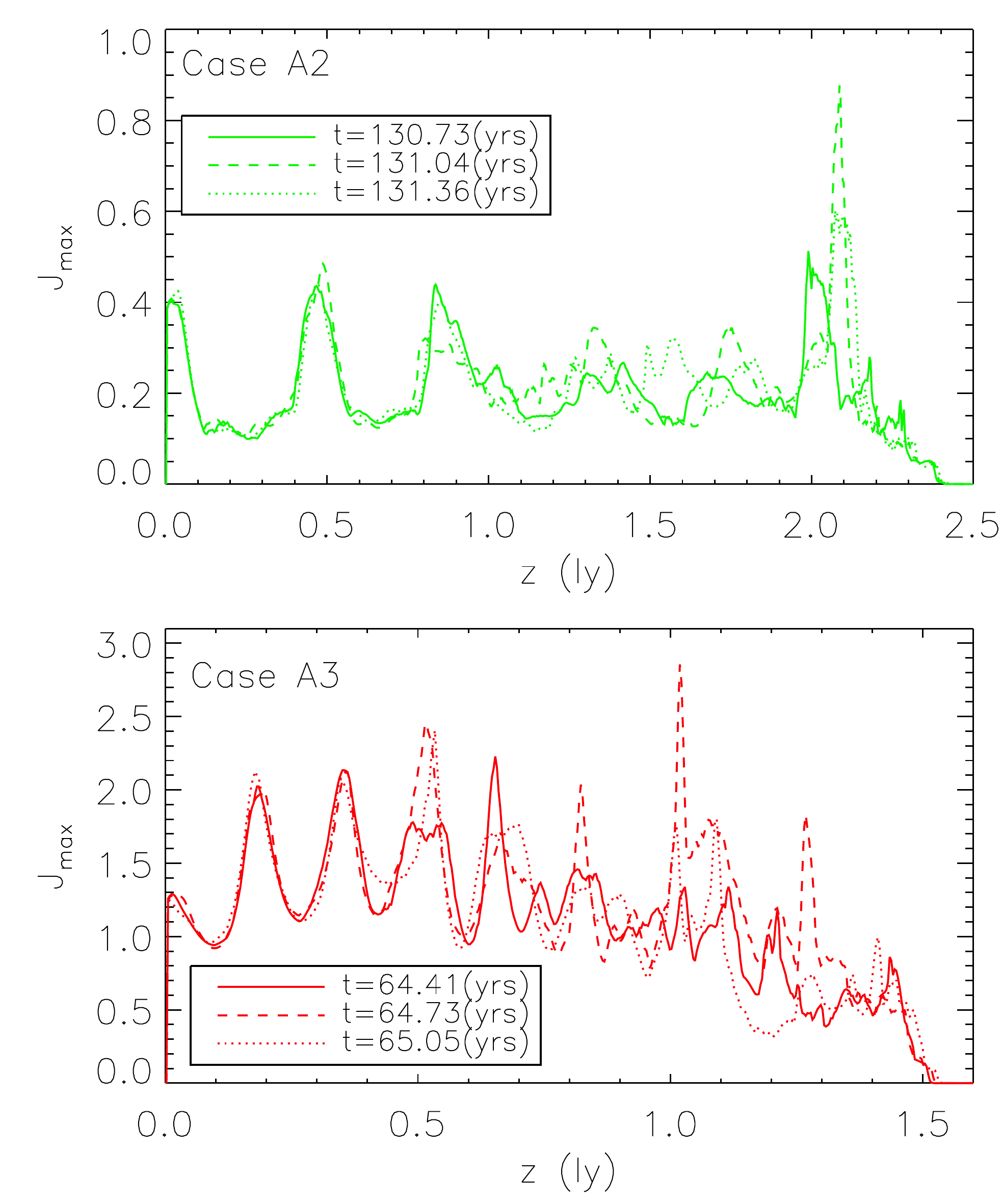}
  \caption{\small Maximum current density taken over $xy$ planes as a function of $z$ for the A2 jet (top) and A3 jet (bottom).
  The solid, dashed and dotted lines mark, respectively, three different close-by simulation times reported in the corresponding legends.
  The largest peaks are shown by the dashed lines.}
  \label{fig:Jmax}
\end{figure}

\begin{figure}
  \includegraphics*[width=0.5\textwidth]{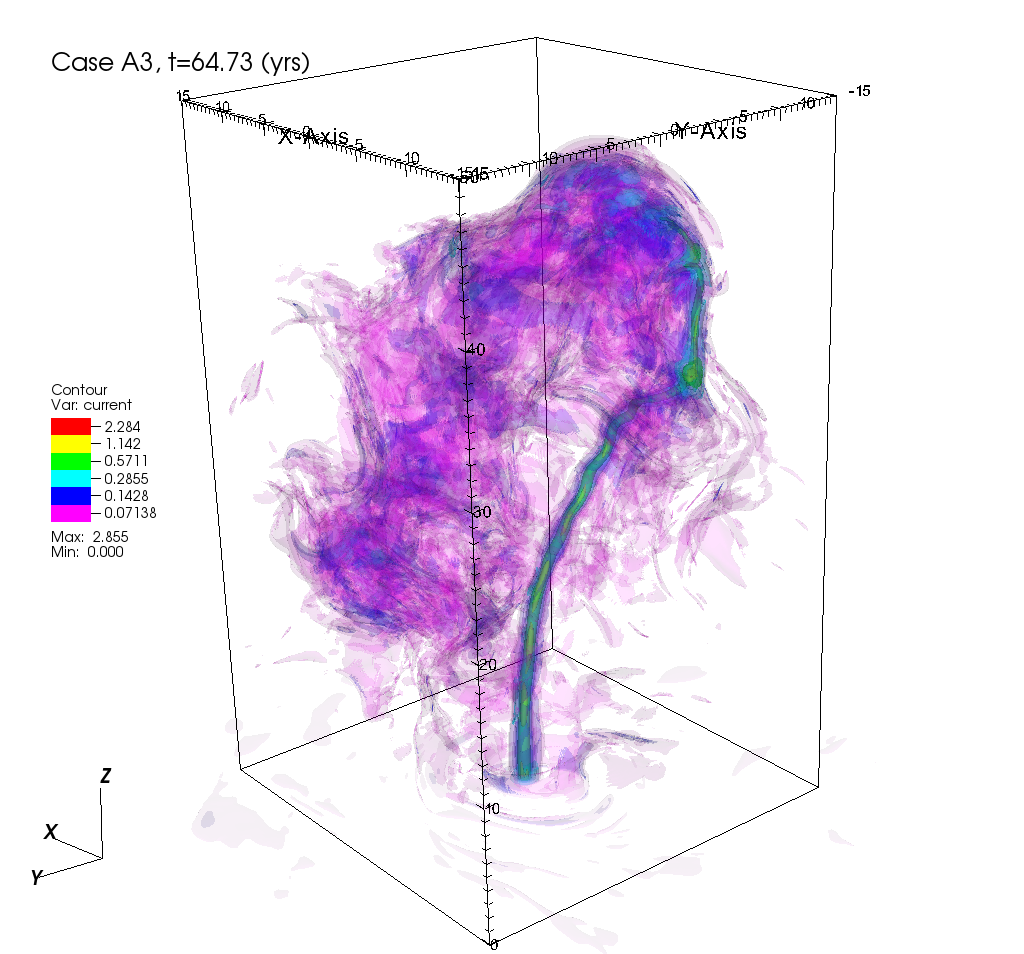}
  \caption{\small Three-dimensional contour plot of the current density after $t=64.73$ years for the A3 jet.
  Six surfaces of constant current density are displayed using different colors, see the attached legend.
  The current peak at $z\approx 1 \,{\rm ly}$ corresponds to the one shown by the dashed line in the bottom panel of Fig. \ref{fig:Jmax}.}
  \label{fig:A3.current.zoom.0204}
\end{figure}

Although our simulations do not include dissipation mechanisms other than numerical that acts at the cell size, it is instructive to localize strong current sheets that may host  regions where particle acceleration is likely to take place.
In Fig. \ref{fig:Jmax} we plot the maximum value of the current density, $J_{\max}(z) = \max_{xy}\left|\nabla\times\vec{B}/(4\pi)\right|$ as a function of the vertical height at three successive simulation times for the A2 and A3 jets.
Current peaks take place in regions of strong pinching where the flow is shocked.
According to the broad distinction of back-end and front-end regions given in Section \ref{sec:internal_struct}, these structures may have considerably different lifetimes.
While the inner regions of the jet have reached a quasi-steady periodic structure with little time-variability, the outer regions reveals the formation of short-lived current peaks that diffuse on a time scale which is of the same order or less than our temporal resolution, approximately $\sim 3.8$ months.
The 3D spatial distribution of the current density, corresponding to the dashed line in the bottom panel of Fig. \ref{fig:Jmax}, is shown at $t=64.73$ yrs in Fig. \ref{fig:A3.current.zoom.0204} for the A3 jet.
Note that the formation of the strong current peak at $z\approx 1 \,{\rm ly}$ occurs concurrently with the development of a jet kink and the abrupt change of flow direction.
The sudden rise of these localized current peaks favors the formation of reconnection layers where induced electric fields may lead to efficient particle acceleration.
This possibility will be addressed in future works.

\section{Summary}
\label{sec:summary}
%
%
%

In this work, we have presented numerical simulations of three-dimensional relativistic magnetized jets propagating into a heavier supernova remnant.
The proposed jet models aim at investigating the dynamics and morphology of the bipolar outflows observed in the Crab Nebula.
Our initial configuration stems from the results of previous 2D axisymmetric numerical models of Pulsar wind nebulae \citep{dZAB.2004} that predict the formation of hot under-dense jets as the result of the magnetic hoop stress collimation process.
Using these models to constrain our input parameters, we have performed numerical simulations of jets initially carrying a purely azimuthal field by varying the bulk flow Lorentz factor $\gamma$ and the ratio between Poynting and kinetic energy fluxes ($\sigma$) at the injection region.

Our results show that jets with moderately/high magnetic fields ($1 \lesssim \sigma \lesssim 10$) are prone to large scale non-axisymmetric current-driven instabilities leading to prominent deflections of the jet beam away from the axis.
This effect is enhanced by the high density contrast ($\sim 10^6$) between the jet and the supernova remnant which leads to the formation of a strongly perturbed beam and largely over-pressurized cocoons.
The typical timescale for the formation of these curved patterns is, for the representative parameters adopted in our model, of the order of a few years and thus compatible with observations.
While the $|m|=1$ kink mode is mainly responsible for the formation of non axisymmetric morphologies, the presence of axial pinch modes engenders a knotty structure with a chain of strong intermediate shocks with large compression factors.
High variability is observed in the outer regions where the dynamics is strongly influenced by the interaction of the jet material with the ambient medium.
Here rapid variations of the flow properties are characterized by the formation of intermediate magnetized shocks in proximity of sudden kinked deflection of the flow trajectory.
These are short-lived episodes leading to the formation of intermittent unstable structures such as sporadic jet fragmentation or strong reconnection layers on a time-scale of a few months.

Our computations demonstrate that the development of these unstable patterns is more pronounced in relatively low-$\gamma$ jets ($\gamma_{j}\approx 2$) which show the largest deflections and evolve entirely inside the remnant.
Conversely, jets with larger Lorentz factor ($\gamma_{j}\gtrsim 4$) propagate with larger inertia, are less affected by the growth of pinch or kink modes and drill out of the remnant in less than $50$ years.
Furthermore, jets with low magnetic fields ($\sigma \approx 0.1$) are weakly affected by the onset of current-driven modes and tend to propagate more parallel to the longitudinal axis.

Based on our relativistic 3D MHD simulations, for the first time we can conclude that moderately to high-$\sigma$ jets with relatively small Lorentz factors (case A2 and A3) are the most likely candidates to account for the dynamical behavior observed in the Crab Nebula jet. In particular, the change in orientation of the jet recently noticed (Weisskopf et al. in preparation) can be reproduced by our simulations. 
Our findings are complementary to 3D numerical calculations \citep{Mizuno.2011, Porth} studying the effects of large values of $\sigma$ in pulsar winds.

Future extensions of this work will take into account the full three-dimensional structure of the PWN allowing us to model the jet launching region and the collimation process. 
Furthermore, our results can be used as a starting point for a detailed analysis of the impulsive particle acceleration mechanism in the Crab Nebula as due to induced electric fields at the localized reconnection layers in the termination zone of the jet.

\section*{Acknowledgments}

We thank Martin Weisskopf for providing Fig. 1 and for many stimulating
discussions about the Crab Nebula morphology and the characteristics of
the South East jet.
A.M. wishes to thank Luca Del Zanna and Elena Amato for valuable comments on the initial jet configuration and G. Bodo for helpful discussions on current-driven instabilities.
We acknowledge the CINECA award under the ISCRA initiative, for the availability of high performance computing resources and support.
This work was partially supported by ASI grants n. I/042/10/0, I/028/12/0.


\label{lastpage}

\end{document}